\documentclass[12pt]{article}
\pdfoutput =1
\usepackage{graphics}
\usepackage{graphicx}
\DeclareGraphicsExtensions{.pdf}
\usepackage{float} %Include figure filesusepackage{graphicx} %Include figure files
\usepackage{subfloat}
\usepackage{amsmath}
\usepackage{slashed}
\usepackage{amsfonts}  
\usepackage{amssymb}

\def\m{\mu}

\usepackage{color}

\usepackage{subcaption}
\begin{document}
\date{}
%%%%%%%%%%%%%%%%%%%%
\title{{\bf{\Large $AdS_2$ holography and ModMax }}}
%%%%%%%%%%%%%%%%%%%%
%\iffalse
\author{
 {\bf {\normalsize Hemant Rathi\thanks{E-mail:  hrathi07@gmail.com, hrathi@ph.iitr.ac.in}~ and ~Dibakar Roychowdhury}$
$\thanks{E-mail:  dibakarphys@gmail.com, dibakar.roychowdhury@ph.iitr.ac.in}}\\
 {\normalsize  Department of Physics, Indian Institute of Technology Roorkee,}\\
  {\normalsize Roorkee 247667, Uttarakhand, India}
\\[0.3cm]
}
%\fi
%\date{}
\maketitle
%%%%%%%%%%%%%%%%%%%%%%%%%%%%%%%%%%%%%%%%%%%%%%%%%%%%%%%%%%%%
\abstract{We present a JT gravity set up in the presence of projected ModMax corrections in two dimensions. Our starting point is the Einstein's gravity in four dimensions accompanied by the ModMax Lagrangian. The 2D gravity action is obtained following a suitable dimensional reduction which contains a 2D image of the 4D ModMax Lagrangian. We carry out a perturbative analysis to find out the vacuum structure of the theory which asymptotes to $AdS_2$ in the absence of $U(1)$ gauge fields. We estimate the holographic central charge and obtain corrections perturbatively upto quadratic order in the ModMax and the $U(1)$ coupling. We also find out ModMax corrected 2D black hole solutions and discuss their extremal limits.  }
%%%%%%%%%%%%%%%%%%%%%%%%%%%%%%%%%%%%
\section{Overview and motivation}

The non-linear generalisations of Maxwell electrodynamics \cite{Born:1934gh}-\cite{Heisenberg:1936nmg} in four dimensions play a pivotal role in understanding the dynamics of charged particles in the strong field regime. For example, the Born-Infeld (BI) theory \cite{Born:1934gh} was proposed in order to obtain the finite self energy corrections for a charged particle in an electromagnetic field. On the other hand, the Heisenberg-Euler-Kockel (HEK) model \cite{Heisenberg:1936nmg} describes the vacuum polarization effects of Quantum Electrodynamics\footnote{See a recent review \cite{Sorokin:2021tge} for different versions of the non-linear modified theories of Maxwell electrodynamics.}. However, both of these (non-linear) theories meet the standard Maxwell electrodynamics in the limit of ``weak'' field approximations.

Generally, the non-linear generalizations of Maxwell electrodynamics (NLE) is characterised by an action that contains a Lorentz scalar and a pseudo scalar which are quadratic in the field strength ($F^{\mu\nu}$) \cite{Sorokin:2021tge}-\cite{Peres:1961zz}
\begin{align}\label{introlinv}
    S= \frac{1}{2}F_{\mu\nu}F^{\mu\nu}\hspace{1mm},\hspace{2mm}P=\frac{1}{2}F_{\mu\nu}\Tilde{F}^{\mu\nu},
\end{align}
where $\Tilde{F}^{\mu\nu}$ is the Hodge dual of $F^{\mu\nu}$.

For instance, the BI electrodynamics is described by the following Lagrangian density \cite{Born:1934gh}
\begin{align}\label{introbi}
    \mathcal{L}_{BI}=T-\sqrt{T^2+\frac{T}{2}F_{\mu\nu}F^{\mu\nu}-\frac{1}{16}\big(F_{\mu\nu}\Tilde{F}^{\mu\nu}\big)^2},
\end{align}
where $T$ is the coupling parameter having the dimension of energy density.  Clearly, in the weak field limit ($T\rightarrow\infty$), the Lagrangian density (\ref{introbi}) reduces to the standard Maxwell electrodynamics. 

Unlike the standard Maxwell electrodynamics, its non-linear modifications are generally not invariant under the $SO(2)$ duality transformations and in fact break the conformal symmetry in four dimensions. For instance, the HEK theory \cite{Heisenberg:1936nmg} is not invariant under the electromagnetic duality and does not have a conformal symmetry. However, the BI electrodynamics is invariant under the $SO(2)$ duality \cite{Gibbons:1995cv} although it is not conformal invariant due to the presence of the dimensionful coupling ($T$) in the theory (\ref{introbi}).

Recently, there has been a radical proposal \cite{Bandos:2020jsw}-\cite{Kosyakov:2020wxv} to (non-linearly) generalize the Maxwell electrodynamics which retains its conformal invariance (in four dimensions) as well as preserves the $SO(2)$ duality symmetry. This goes under the name of the ``ModMax'' electrodynamics\footnote{For details, see the recent review \cite{Sorokin:2021tge}.}.  

The ModMax electrodynamics is a 1-parameter deformation of the Maxwell electrodynamics in four dimensions that is described by the following Lagrangian density\footnote{In the limit $\gamma\rightarrow0$, the ModMax electrodynamics  reduces to the standard Maxwell electrodynamics.} \cite{Bandos:2020jsw}-\cite{Kosyakov:2020wxv} 
\begin{align}\label{introld}
     \mathcal{L}_{MM}&=\frac{1}{2}\Big(S \cosh{\gamma}-\sqrt{S^2+P^2}\sinh{\gamma}\Big),
\end{align}
where $\gamma$ is the dimensionless coupling constant that measures the strength of the electromagnetic self interaction.

The physical requirements that the theory must be unitary and preserves the causality restrict the ModMax parameter $(\gamma)$ to take only positive values ($\gamma>0$)  \cite{Bandos:2020jsw}. The above restriction guarantees that the Lagrangian density (\ref{introld}) is a convex function of the electric field strength $E^i$.  

There have been some further modifications to the ModMax electrodynamics in the literature which include the 1-parameter generalisation of the BI theory\footnote{In the weak field limit, the ($\gamma$BI) theory reduces to the standard ModMax electrodynamics (\ref{introld}).} ($\gamma$BI) \cite{Bandos:2020hgy}  and $\mathcal{N}=1$ supersymmetric extension of the ModMax electrodynamics\footnote{See \cite{Kruglov:2021bhs}-\cite{ Cano:2021tfs} for further details. }  \cite{Bandos:2021rqy}. The supersymmetric version of the ModMax electrodynamics is invariant under the electromagnetic duality as well as posses the superconformal symmetry \cite{Bandos:2021rqy}.

The ModMax electrodynamics finds an extensive application in theories of gravity \cite{Barrientos:2022bzm}-\cite{Amirabi:2020mzv} as well. In fact, a large number of solutions have been obtained down the line. For instance, accelerated black holes \cite{Barrientos:2022bzm}, the Taub-NUT \cite{BallonBordo:2020jtw}-\cite{Flores-Alfonso:2020nnd} and Reissner-Nordstorm solutions \cite{Amirabi:2020mzv} in diverse spacetime dimensions have been constructed in the presence of ModMax interactions and the effects of non-linearity were explored on their thermal properties. Recently, the non-linear models of electrodynamics have also found their applications in the context of strongly correlated systems  \cite{Cai:2008in}-\cite{Hartnoll:2008vx} by means of the celebrated $AdS_{d+1}/CFT_d$ correspondence  \cite{Maldacena:1997re}-\cite{Gubser:1998bc}.

Despite of several notable applications those are alluded to the above, ModMax theories are least explored in $ AdS_2$ holography and in particular in the context of the JT/SYK correspondence \cite{Jackiw:1984je}-\cite{Gross:2016kjj}. The purpose of the present paper is to fill up some of these gaps in the literature and find out an interpretation for the projected ModMax interactions within the realm of 2D gravity theories.

The pure Jackiw-Teitelboim (JT) gravity \cite{Jackiw:1984je}-\cite{Teitelboim:1983ux} is the two dimensional theory of Einstein-dilaton gravity in the presence of a negative cosmological constant. Under certain special circumstances, this theory is conjectured to be the dual description of the Sachdev-Ye-Kitaev (SYK) model \cite{Sachdev:1992fk}-\cite{Gross:2016kjj} which  is a quantum mechanical theory of N interacting (Majorana) fermions in one dimension\footnote{The bulk dual of the pure SYK model contains an infinite tower of massive particles which are dual to primary $O(N)$ singlet operators \cite{Polchinski:2016xgd}, \cite{Gross:2017hcz}. However, the pure JT gravity does not contain the tower of such massive particles. Therefore, the SYK/JT correspondence should make sense only in the soft/Schwarzian limit.} . Interestingly, this model can be solved exactly at strong coupling and in the Large N limit. The generalisation of the JT/SYK correspondence in the presence of $U(1)$ gauge fields and SU(2) Yang-Mills fields have been carried out in a series of papers \cite{Davison:2016ngz}-\cite{Lala:2020lge}. 

In the present paper, we cook up a theory of JT gravity in the presence of 2D ``projected'' ModMax interactions and compute various physical entities associated with the boundary theory. For instance, we construct the holographic stress-energy tensor \cite{Rathi:2021aaw}, \cite{Castro:2008ms}, \cite{Cadoni:2000gm}-\cite{Narayan:2020pyj} and compute the associated central charge \cite{Rathi:2021aaw}, \cite{Castro:2008ms}, \cite{Cadoni:2000gm}, \cite{Hartman:2008dq} for the boundary theory. Finally, we construct black hole solutions in two dimensions and explore the effects of projected ModMax interactions on their thermal behaviour.

The organisation for the rest of the paper is as follows :

$\bullet$ In Section \ref{secmd2d}, we follow suitable dimensional reduction procedure \cite{Davison:2016ngz}-\cite{Rathi:2021aaw}, \cite{Lala:2020lge} to construct a model for JT gravity in the presence of 2D projected ModMax interactions. We also clarify the meaning of projected ModMax interactions in 2D and in particular present a detail comparison with the 4D ModMax interactions.

$\bullet$ In Section \ref{secsol}, we calculate the conformal dimensions of different scalar operator in deep IR limit and make a comparative analysis between them. We further explore the vacuum structure of the theory using the Fefferman-Graham gauge \cite{Rathi:2021aaw}, \cite{fefferman} by treating the non-linear $U(1)$ gauge interactions as ``perturbations'' over the pure JT gravity solutions. We estimate these solutions upto quadratic order in the gauge and ModMax couplings.

$\bullet$ In Section \ref{secstcc}, we construct the ``renormalised'' boundary stress tensor and investigate its transformation properties under the combined action of the diffeomorphism and the $U(1)$ gauge transformations  \cite{Castro:2008ms}. {We compute the central charge $(c_M)$ associated with the boundary theory  \cite{Castro:2008ms} up to quadratic order in the (ModMax and $U(1)$) couplings.

$\bullet$ In Section \ref{secbhwithmm}, we construct the black hole solutions upto quadratic order in the couplings. We observe that the non-linear interactions (or the projected ModMax interactions) play a crucial role in obtaining a finite value for the background fields at the horizon. 

Furthermore, we compute the Hawking temperature for 2D black holes \cite{Hawking:1975vcx} and calculate the associated Wald entropy \cite{Wald:1993nt}-\cite{Pedraza:2021cvx}. We also investigate the ``extremal'' limit associated with these 2D black hole solutions and calculate the corresponding Wald entropy.

$\bullet$ We draw our conclusion in Section \ref{secconc}, along with some future remarks.

%%%%%%%%%%%%%%%%%%%%%%%%%%%%%%%%%%%
\section{JT gravity and 2D projected ModMax }\label{secmd2d}
The ModMax theory coupled to Einstein gravity in four dimensions is defined as \cite{Bandos:2020jsw}-\cite{Kosyakov:2020wxv}, \cite{Barrientos:2022bzm}
\begin{align}\label{lmd4d}
    I^{(4)}=\frac{1}{16\pi G_4}\int d^4x\sqrt{-g_{(4)}}\Big(R^{(4)}-2\Lambda -4\kappa\mathcal{L}^{(4)}_{\text{MM}}\Big),
\end{align}
where $R^{(4)}$ is the Ricci scalar in 4 dimensions, $\Lambda=-3$ is the cosmological constant\footnote{Here, we set the $AdS$ length $l=1$.}, $G_4$ is the Newton's constant in four dimensions,  $\kappa$ is the coupling constant and $\mathcal{L}^{(4)}_{\text{MM}}$ is the ModMax Lagrangian density in four dimensions \cite{Bandos:2020jsw}-\cite{Kosyakov:2020wxv}, \cite{Barrientos:2022bzm}

\begin{align}\label{momdmaxld}
    \mathcal{L}^{(4)}_{\text{MM}}&=\frac{1}{2}\Big(S \cosh{\gamma}-\sqrt{S^2+P^2}\sinh{\gamma}\Big),\nonumber\\
    \hspace{2mm}S&=\frac{1}{2}F_{MN}F^{MN}\hspace{1mm},\hspace{2mm}P=\frac{1}{2}F_{MN}\Tilde{F}^{MN}\hspace{1mm}, \hspace{2mm}\Tilde{F}^{MN}=\frac{1}{2}\epsilon^{MNUV}F_{UV}.
\end{align}

Here, $\gamma$ is the ModMax parameter and $(M,N)$ are the 4 dimensional space-time indices. Clearly, the standard Maxwell electrodynamics is recovered in the limit $\gamma\rightarrow 0$  \cite{Bandos:2020jsw}-\cite{Kosyakov:2020wxv}, \cite{Barrientos:2022bzm}.

The imprint of the ModMax theory (\ref{momdmaxld}) in two dimensions can be obtained via dimensional reduction \cite{Davison:2016ngz}-\cite{Rathi:2021aaw}, \cite{Lala:2020lge} of the following form
\begin{align}\label{ansatz}
    ds_{(4)}^2&=ds_{(2)}^2+\Phi(x^{\mu}) dx_i^2\hspace{1mm},\hspace{2mm}ds_{(2)}^2=g_{\mu\nu}(x^{\alpha})dx^{\mu}dx^{\nu},\nonumber\\
    A_{\mu}&\equiv A_{\mu}(x^{\nu}),\hspace{1mm}A_i\equiv A_{i}(x^{\mu}),
\end{align}
where $(\mu,\nu)$ are the two dimensional indices and $(i,j)$ are the indices of the compact dimensions.

Substituting (\ref{ansatz}) into (\ref{lmd4d}) and integrating over the compact directions, one finds\footnote{The Newton's constant in two and four dimensions are related by $G_2=\frac{G_4}{V_2}$, where $V_2$ is the volume of the compact space.}
\begin{align}\label{lmd2d}
  I_{\text{bulk}}=\frac{1}{16\pi G_2}\int d^2x\sqrt{-g_{(2)}}\Big(\Phi R^{(2)}-2\Lambda\Phi-4\kappa\Phi\mathcal{L}^{(2)}_{\text(MM)}\Big), 
\end{align}
where $R^{(2)}$ is the Ricci scalar in two dimensions, $G_2$ is the Newton's constant in two dimensions and 
\begin{align}\label{onlylmm2d}
    \mathcal{L}^{(2)}_{\text{MM}}&=\frac{1}{2}\Big(s\cosh{\gamma}-\sqrt{s^2+p^2}\sinh{\gamma}\Big),\nonumber\\
    s&=\frac{1}{2}F_{\mu\nu}F^{\mu\nu}+\Phi^{-1}\Big((\partial \chi)^2+(\partial \xi)^2\Big),\hspace{1mm}p=-2\Phi^{-1}\epsilon^{\mu\nu}\partial_{\mu}\chi\partial_{\nu}\xi
\end{align}
is what we define as the Lagrangian density of the projected ModMax theory in two dimensions. Here, we denote $A_{2}=\chi(x^{\mu}),\hspace{1mm} A_{3}=\xi(x^{\mu})$ and introduce $\epsilon^{\mu\nu}=\frac{\varepsilon^{\mu\nu}}{\sqrt{-g_{(2)}}}$ as the Levi-Civita tensor in two dimensions. 

Notice that, in the limit $\gamma\rightarrow0$, we do not recover the standard Maxwell electrodynamics in two dimensions \cite{Rathi:2021aaw}, \cite{Castro:2008ms}-\cite{Lala:2020lge}. On contrary, we do have additional contributions coming from non-vanishing scalar fields $\xi$ and $\chi$ which arise by virtue of the dimensional reduction procedure. This turns out to be the unique feature of the projected ModMax interactions in two dimensions. The $\gamma\rightarrow 0$ limit is what we refer as the 2D Maxwell interaction in this paper.

$\bullet$ \textbf{A comparative study of 4D ModMax and the 2D projected ModMax:}

Below, we draw a comparative analysis between 4D ModMax \cite{Bandos:2020jsw}-\cite{ Kosyakov:2020wxv} and its 2D projection which plays the central role in what follows. 4D ModMax preserves the conformal invariance in its usual sense which is also evident from the generic structure of the associated stress-energy tensor 
 
\begin{align}\label{tr4}
    T^{(4)}_{MN}\sim f(\gamma)\Bigg(-\frac{1}{2}F^2g_{MN}+2g^{QP}F_{QM}F_{PN}\Bigg),
\end{align}
where we define the function
\begin{align}
   f(\gamma)= \Bigg(\cosh{\gamma}-\frac{F^2\sinh{\gamma}}{\sqrt{\big(F_{RS}F^{RS}\big)^2+\big(F_{RS}\Tilde{F}^{RS}\big)^2}}\Bigg).
\end{align}

Clearly, the trace $T^{M(4)}_M$ vanishes identically in four dimensions. On the other hand, the trace of the projected ModMax in two dimensions turns out to be
\begin{align}
   T^{\mu(2)}_{\mu}= g^{\mu\nu}T_{\mu\nu}^{(2)}= \frac{\Phi F^2}{2}\Bigg(\cosh{\gamma}-\frac{s\sinh{\gamma}}{\sqrt{s^2+p^2}}\Bigg),
\end{align}
which is a non-vanishing entity. 

This reflects to the fact that the projected theory losses its conformal invariance in two dimensions. Furthermore, the absence of the (Hodge) dual two form ($\Tilde{F}^{\mu\nu}$) in two dimensions spoils the electromagentic SO(2) duality invariance of the 2D projected theory in comparison to its 4D cousin. However, it is noteworthy to mention that the ModMax coupling $(\gamma)$ that appears in the 2D projected version is same as that of the 4D parent theory.

The equations of motion corresponding to different field contents can be obtained by varying the action (\ref{lmd2d}) 

\begin{align}\label{bulkvar}
        \delta I_{\text{bulk}}= \frac{1}{16\pi G_2}\int d^2x\sqrt{-g}\Big(\mathcal{H}_{\mu\nu}\delta g^{\mu\nu}+\mathcal{H}_{\Phi}\delta \Phi+\mathcal{H}_{\mu}\delta A^{\mu}+\mathcal{H}_{\chi}\delta\chi+\mathcal{H}_{\xi}\delta\xi\Big),
    \end{align}
where we define individual entities as 
\begin{align}
\mathcal{H}_{\Phi}=\hspace{1mm}&R-2\Lambda-4\kappa\mathcal{L}^{(2)}_{\text{MM}}+2\kappa\Phi^{-1}\Bigg[\Big((\partial \xi)^2+(\partial \chi)^2\Big)\cosh{\gamma}-\Bigg\{\frac{s\Big((\partial \xi)^2+(\partial \chi)^2\Big)}{\sqrt{s^2+p^2}}-\nonumber\\
    & \frac{2p\epsilon^{\mu\nu}\nabla_{\mu}\chi\nabla_{\nu}\xi}{\sqrt{s^2+p^2}}\Bigg\}\sinh{\gamma}\Bigg]=0,\label{geneomi}\\
\mathcal{H}_{\mu\nu}=\hspace{1mm}& \square \Phi g_{\mu\nu}-\nabla _{\mu}\nabla_{\nu}\Phi+\Lambda\Phi g_{\mu\nu} -2\kappa\Phi\Bigg[\mathcal{F_{\mu\nu}}\cosh{\gamma}-\frac{\sinh{\gamma}}{\sqrt{s^2+p^2}}\Bigg(s\mathcal{F_{\mu\nu}}-\frac{1}{2}s^2g_{\mu\nu}\Bigg)\nonumber\\&-\frac{s}{2}g_{\mu\nu}\cosh{\gamma}\Bigg]=0,\label{geneomijhi}\\
    \mathcal{H}_{\chi}=\hspace{1mm}&\kappa\nabla_{\mu}\Bigg[\nabla^{\mu}\chi \cosh{\gamma}-\frac{s\nabla^{\mu}\chi-p\epsilon^{\mu\nu}\nabla_{\nu}\xi}{\sqrt{s^2+p^2}}\sinh{\gamma}\Bigg]=0,\\
     \mathcal{H}_{\xi}=\hspace{1mm}&\kappa\nabla_{\mu}\Bigg[\nabla^{\mu}\xi \cosh{\gamma}-\frac{s\nabla^{\mu}\xi+p\epsilon^{\mu\nu}\nabla_{\nu}\chi}{\sqrt{s^2+p^2}}\sinh{\gamma}\Bigg]=0,\label{geneomijhf}\\
      \mathcal{H}_{\mu}=\hspace{1mm}&\kappa\nabla_{\mu}\Bigg[\Phi \Bigg(\cosh{\gamma}-\frac{s\sinh{\gamma}}{\sqrt{s^2+p^2}}\Bigg)F^{\mu\nu}\Bigg]=0,\label{geneomf}
\end{align}
along with the function
    \begin{align}
   \mathcal{F_{\mu\nu}}=&\hspace{1mm}F_{\mu\alpha}F_{\nu\beta}g^{\alpha\beta}+\Phi^{-1}\Big(\partial_{\mu}\xi\partial_{\nu}\xi+\partial_{\mu}\chi\partial_{\nu}\chi\Big).
   \end{align}
   
 %%%%%%%%%%%%%%%%%%%%%%%%%%%%%%%%%%%%%%%%%%%%%%%%%%%
    
 \section{General solution with 2D projected ModMax}\label{secsol}
 The purpose of this Section is to obtain the most general solutions of (\ref{geneomi})-(\ref{geneomf}) in the Fefferman-Graham gauge\footnote{The explicit form of these equations (\ref{geneomi})-(\ref{geneomf}) have been provided in the Appendix \ref{geneomap}.} \cite{Rathi:2021aaw}, \cite{fefferman}
 \begin{align}\label{fggauge}
       ds^2=\hspace{1mm}&d\eta^2+h_{tt}(t,\eta)dt^2,\hspace{1mm}A_{\mu}dx^{\mu}=\hspace{1mm}A_t(t,\eta)dt,\nonumber\\
       \Phi=\hspace{1mm}&\Phi(t,\eta),\hspace{1mm}\chi=\chi(t,\eta),\hspace{1mm}\xi=\xi(t,\eta).
 \end{align}

$\bullet$ \textbf{A note on conformal dimensions:}

 Here, we present a calculation on the conformal dimensions of the dual operators $\Delta_{\chi}$, $\Delta_{\xi}$ and $\Delta_{\Phi}$ corresponding to the bulk scalar fields $\chi$, $\xi$ and $\Phi$ respectively. This allows us to make a comparative study between various operator dimensions in the deep IR limit.

 The IR fixed point \cite{Castro:2018ffi}-\cite{Castro:2021wzn} is defined as the set of solutions to the equations of motion (\ref{geneomi})-(\ref{geneomf}) for constant values of the scalar fields
 \begin{align}\label{irjfp}
\chi(t,\eta)=\chi^{*}\hspace{1mm},\hspace{2mm}\xi(t,\eta)=\xi^{*}\hspace{1mm},\hspace{2mm} \Phi(t,\eta)=\Phi^{*},
 \end{align}
 where the superscript `*' denotes the values of the background scalars at the IR fixed point.

Using (\ref{irjfp}), one can solve the above set of equations (\ref{geneomi})-(\ref{geneomf}) in the Fefferman-Graham gauge (\ref{fggauge}) to obtain 
 \begin{align}
\omega^*=&\hspace{1mm}\alpha(t)e^{\sqrt{2}\eta\lambda}+\beta(t)e^{-\sqrt{2}\eta\lambda},\\
     A_{t}^*=&\hspace{1mm}\mu(t)+\frac{c}{\sqrt{2}\lambda}\Big(\alpha(t)e^{\sqrt{2}\eta\lambda}-\beta(t)e^{-\sqrt{2}\eta\lambda}\Big),
 \end{align}
where we define  $\lambda=\sqrt{-\Lambda}=\sqrt{3}$, $\omega=\sqrt{-h_{tt}}$ and $c$ is the integration constant. Here, $\alpha(t)$, $\beta(t)$ and $\mu(t)$ are some arbitrary functions of time.

 In order to compute the conformal dimensions of the dual operators, we expand the scalar fields ($\chi$, $\xi$ and $\Phi$) around the fixed point (\ref{irjfp}) and retain the equations of motion (\ref{geneomijhi})-(\ref{geneomijhf}) upto linear order in scalar fluctuations which yields 

 \begin{align}
     \Bigg[\partial^2_{\eta}+\frac{1}{\omega^*}\Big(\partial_{\eta}\omega^*\Big)\partial_{\eta}-\frac{1}{\omega^*}\partial_t\Bigg(\frac{1}{\omega^*}\partial_t\Bigg)-m^2\Bigg]\Tilde{\Phi}=&\hspace{1mm}0,\label{pertj1}\\
     \Bigg[\partial_{\eta}\Big(\omega^*\partial_{\eta}\Big)-\partial_t\Bigg(\frac{1}{\omega^*}\partial_t\Bigg)\Bigg]\Tilde{\chi}=&\hspace{1mm}0,\label{pertj2}\\
     \Bigg[\partial_{\eta}\Big(\omega^*\partial_{\eta}\Big)-\partial_t\Bigg(\frac{1}{\omega^*}\partial_t\Bigg)\Bigg]\Tilde{\xi}=&\hspace{1mm}0,\label{pertj3}
 \end{align}
 where we define $m^2=\big(6-2c^2\kappa e^{-\gamma}\big)$ and scalar fluctuations $\Tilde{\mathcal{Y}}=\mathcal{Y}-\mathcal{Y}^*$, where $\mathcal{Y}$ collectively denotes the scalar fields ($\Phi$, $\chi$ and $\xi$).
 
 It should be noted that, the mass-squared term ($m^2$) defined above must satisfy the Breitenlohner-Freedman (BF) bound\footnote{In ($d+1$) spacetime dimensions, the BF bound is defined as $m^2\geq-\Big(\frac{d^2}{2L}\Big)^2,$ where $L$ is the AdS length.} \cite{Ramallo:2013bua}, which for the present example sets a constraint of the form $c\leq\sqrt{\frac{25e^{\gamma}}{8\kappa}}$. Notice that, unlike (\ref{pertj1}), the equations of motion for scalar fluctuations $\Tilde{\chi}$ (\ref{pertj2}) and $\Tilde{\xi}$ (\ref{pertj3}) do not contain any mass-squared term. This indicates that these scalar fields ($\chi$ and $\xi$) are massless. This is consistent with the fact that these scalar fields ($\chi$ and $\xi$) carry only kinetic terms in the Lagrangian (\ref{onlylmm2d}).

 From the above set of equations (\ref{pertj1})-(\ref{pertj3}), one could finally decode the conformal dimensions\footnote{The conformal dimension of the dual operator ($\Delta$) is defined as $\Delta(\Delta-1)=m^2$ \cite{Garcia-Garcia:2020ttf}, \cite{Castro:2018ffi}, \cite{Ramallo:2013bua}, where $m$ represents the mass of the scalar field.} of the dual operators as 
 \begin{align}\label{jhepsop}
     \Delta_{\Phi\pm}=\hspace{1mm}\frac{1}{2} \left(1\pm\sqrt{25-8 e^{-\gamma } c^2 \kappa }\right)\hspace{1mm},\hspace{2mm}\Delta_{\chi}=\hspace{1mm}\Delta_{\xi}=\hspace{1mm}1,
 \end{align}
where the subscript `$\pm$' denotes the two possible values of $\Delta_{\Phi}$.  
 
It is interesting to notice that the conformal dimension, $(\Delta_{\chi}=\Delta_{\xi})>\Delta_{\Phi+}$ for the range of constant, $\sqrt{\frac{3e^{\gamma}}{\kappa}}<c\leq\sqrt{\frac{25e^{\gamma}}{8\kappa}}$. This indicates that the dynamics of the dilaton fluctuation $\Tilde{\Phi}$ dominates \cite{Castro:2018ffi}--\cite{Castro:2021wzn} over the scalar fluctuations $\Tilde{\chi}$ and $\Tilde{\xi}$ in the deep IR. 
 
 On the other hand, one could set the conformal dimension, $(\Delta_{\chi}=\Delta_{\xi})<\Delta_{\Phi+}$ given the range $0\leq c<\sqrt{\frac{3e^{\gamma}}{\kappa}}$, which suggests that the IR dynamics is dominated by the scalar fluctuation $\Tilde{\chi}$ and $\Tilde{\xi}$. However,  for a particular choice of constant $c=\sqrt{\frac{3e^{\gamma}}{\kappa}}$, the  conformal dimensions, $\Delta_{\chi}=\Delta_{\xi}=\Delta_{\Phi+}=1$. In this case, the dynamics of all scalar fluctuations $\Tilde{\mathcal{Y}}$ are equally important in the deep IR.  

On a similar note, one finds that the maximum value of the conformal dimension\footnote{In this case, the constant $c$ is restricted to the range $\sqrt{\frac{3e^{\gamma}}{\kappa}}\leq c\leq\sqrt{\frac{25e^{\gamma}}{8\kappa}}$. If $c<\sqrt{\frac{3e^{\gamma}}{\kappa}}$, then the conformal dimension $\Delta_{\Phi-}$ become negative.} $\Delta_{\Phi-}$ is $1/2$. Therefore, in this case, the dilaton fluctuation $(\Tilde{\Phi})$ always dominates over the scalar fluctuations. Therefore, to summarise, one could conjecture that the dilaton fluctuation always dominates over scalar fluctuation if the constant falls in the range $\sqrt{\frac{3e^{\gamma}}{\kappa}}<c\leq\sqrt{\frac{25e^{\gamma}}{8\kappa}}$. 

Finally, it is noteworthy to compare our results with the existing literature \cite{Castro:2018ffi}-\cite{Castro:2021fhc}. The authors in \cite{Castro:2018ffi}, construct a 2D theory of gravity in the presence of a dilaton ($e^{-2\psi}$), scalar field ($\chi$) and a $U(1)$ gauge field following a consistent reduction of Einstein gravity in five dimensions. Unlike the present example, the authors in \cite{Castro:2018ffi} obtained a mass-squared term for the scalar field ($\chi$) which is thereby used to calculate the conformal dimension of the dual operator. Interestingly, they found that the dual operator is always irrelevant compared to the dilaton operator in the IR. In other words, the dilaton fluctuation  always dominates over the scalar fluctuations in the deep IR. \\

$\bullet$ \textbf{Remarks about perturbative solutions:}
 
 Now, we compute the most general solutions of (\ref{geneomi})-(\ref{geneomf}) in the Fefferman-Graham gauge. Clearly, these  equations (\ref{geneomi})-(\ref{geneomf}) are quite difficult to solve exactly in the ModMax coupling $\gamma$. Therefore, to proceed further, we simplify the fields as $\Phi\equiv\Phi(\eta),\hspace{1mm}h_{tt}\equiv h_{tt}(\eta),\hspace{1mm}A_t\equiv A_t(\eta),\hspace{1mm}\xi\equiv\xi(\eta)\hspace{1mm}\text{and}\hspace{1mm}\chi\equiv\chi(t) $ and solve them ``perturbatively'' treating the 2D Maxwell coupling ($\kappa$) and the 2D ModMax coupling $(\gamma)$ as expansion parameters.

One can systematically expand these fields using the expansion parameters ($\kappa$ and $\gamma$) as
\begin{align}
   & \mathcal{A}=\mathcal{A}_0+\kappa\mathcal{A}_1+\gamma\kappa\mathcal{A}_2+\kappa^2\mathcal{A}_3+...,\label{Afieldexp}\\
   & \mathcal{B}=\mathcal{B}_1+\gamma\mathcal{B}_2+\kappa\mathcal{B}_3+...,\hspace{2mm}|\kappa|<<1,\hspace{1.5mm}|\gamma|<<1,\label{Bfieldexp}
\end{align}
where $\mathcal{A}$ collectively denotes the fields ($\Phi$, $\omega$) and $\mathcal{B}$ denotes the remaining fields ($A_t$, $\chi$, $\xi$). Here, the subscript `0' denotes the pure JT gravity solution. On the other hand, subscripts `$1$' and `$2$' denote the leading order corrections due to the 
 2D Maxwell term and the 2D projected ModMax interaction respectively.  Furthermore, the subscript `$3$' stands for the quadratic order corrections due to the 2D Maxwell term alone.

Notice that, the $\mathcal{B}$ fields (\ref{Bfieldexp}) are expanded differently from that of the $\mathcal{A}$ fields (\ref{Afieldexp}). This is due to the fact that the $\mathcal{B}$ fields are coupled with an overall  2D Maxwell coefficient, $\kappa$ in the Lagrangian (\ref{lmd2d}). Therefore, one should think of the expansion (\ref{Bfieldexp}) to be multiplied with an overall factor of $\kappa$. On the other hand, the effects of the 2D projected ModMax comes into the picture at the quadratic level ($\gamma\kappa$). To summarise, we solve the equations of motion  (\ref{geneomi})-(\ref{geneomf}) up to quadratic order ($\gamma\kappa$ and $\kappa^2$) in the couplings and ignore all the higher order corrections.

\subsection{Zeroth order solution}
In order to obtain the pure JT gravity solutions, one has to take the limits $\kappa\rightarrow0$ and $\gamma\rightarrow0$ in the equations (\ref{geneomi})-(\ref{geneomf}), which yields
\begin{align}
\omega_0''+\Lambda\omega_0=&\hspace{1mm}0,\label{zeroeomi}\\
    \Phi_0''+\Lambda\Phi_0=&\hspace{1mm}0,\\
    \frac{\Phi_0'\omega_0'}{\omega_0}+\Lambda\Phi_0=&\hspace{1mm}0.\label{zeroeomf}
\end{align}

On solving (\ref{zeroeomi})-(\ref{zeroeomf}), one finds
\begin{align}
\omega_0=&\hspace{1mm}a_1 e^{\eta  \lambda }+a_2 e^{-\eta  \lambda },\label{zerosoli}\\
    \Phi_0=&\hspace{1mm}\frac{b_1 }{a_1 \lambda }e^{-\eta  \lambda } \left(a_1 e^{2 \eta  \lambda }-a_2\right),\label{zerosolf}
    \end{align}
    where $a_1,a_2$ and $b_1$ are the integration constants. 
    
    Equations (\ref{zerosoli})-(\ref{zerosolf}) are the zeroth order solutions of the theory (\ref{lmd2d}). In the following Sections, we will be using these solutions to obtain the next to leading order corrections for $\mathcal{A}$ and $\mathcal{B}$.
    
\subsection{ Order $\kappa$ solution}
The leading order corrections to the fields $\mathcal{A}$ and $\mathcal{B}$ are due to the presence of the Maxwell interactions in (\ref{lmd2d}),
\begin{align}\label{ONLYMAXWELLACT}
    \mathcal{L}_{\text{Maxwell}}=\frac{1}{4}F_{\mu\nu}F^{\mu\nu}+\frac{1}{2}\Phi^{-1}\Big((\partial \chi)^2+(\partial \xi)^2\Big).
\end{align}

On comparing the coefficients of $\kappa$ in the equations (\ref{geneomi})-(\ref{geneomf}), we obtain 
\begin{align}
 \omega_0\Phi_1''+\omega_1\Phi_0''-\Big(\Phi_0'\omega_1'+\Phi_1'\omega_0'\Big)+2\omega_0\Bigg(\xi_1'^2+\frac{\dot{\chi_1}^2}{\omega_0^2}\Bigg)=&\hspace{1mm}0,\label{oneeqi}\\
 \omega_1''+\Lambda\omega_1-\frac{A_{t1}'^2}{\omega_0}=&\hspace{1mm}0,\\
 \partial_{\eta}\Bigg(\frac{\Phi_0}{\omega_0}A_{t1}'\Bigg)=&\hspace{1mm}0,\\
    \partial_{\eta}\Big(\omega_0\xi_1'\Big)=&\hspace{1mm}0,\\
    \ddot{\chi}_1=&\hspace{1mm}0. \label{oneeqf}
    \end{align}
   
Using the zeroth order solutions (\ref{zerosoli})-(\ref{zerosolf}), one can solve the above set of equations to yield
    \begin{align}       
    \Phi_1=&\hspace{1mm}\frac{e^{-\eta  \lambda } }{4 \lambda ^2}\Bigg(\frac{4 \lambda }{a_1} \left(a_3 b_1 e^{2 \eta  \lambda }+a_2 \log \left(a_2-a_1 e^{2 \eta
    \lambda }\right)\right)+4 \lambda  e^{2 \eta  \lambda } \Big(2 \eta  \lambda -\nonumber\\
    &\log \left(a_2-a_1 e^{2
   \eta  \lambda }\right)\Big)+\tan ^{-1}\left(\frac{\sqrt{a_1} e^{\eta  \lambda
   }}{\sqrt{a_2}}\right)\Bigg(\frac{1}{a_1^{3/2} \sqrt{a_2}}-\frac{e^{2 \eta  \lambda } }{\sqrt{a_1} a_2^{3/2}}\Bigg)\Bigg),\label{onesoli}\\
   \omega_1=&\hspace{1mm}\frac{c_1^2 }{4 a_2}e^{-\eta  \lambda } \left(2 \eta  \lambda  e^{2 \eta  \lambda }-\frac{1}{a_1}\left(a_1 e^{2 \eta  \lambda
   }+a_2\right) \log \left(a_2-a_1 e^{2 \eta  \lambda }\right)\right)+a_3 e^{\eta  \lambda },\\
    A_{t1}=&\hspace{1mm}c_1 \Bigg[\log \left(a_2-a_1 e^{2 \eta  \lambda }\right)- \eta  \lambda\Bigg]+c_2, \\
   \xi_1=&\hspace{1mm}\frac{e_1 }{\lambda }\tan ^{-1}\left(\frac{\sqrt{a_1} e^{\eta  \lambda }}{\sqrt{a_2}}\right)+e_2,\\
    \chi_1=&\hspace{1mm}d_1t+d_2,\label{onesolf}
    \end{align}
    where $a_3,c_i,d_i$ and $e_i$, $(i=1,2)$ are the integration constants.
    
    Equations (\ref{onesoli})-(\ref{onesolf}) represent the leading order corrections to the fields $\mathcal{A}$ and $\mathcal{B}$ in the presence of the 2D Maxwell interactions (\ref{ONLYMAXWELLACT}).
    
    \subsection{ Order $\gamma\kappa$ solution}
    
    Next, we take into account the projected ModMax interactions and their imprint on the background fields $\mathcal{A}$ (\ref{Afieldexp}) and $\mathcal{B}$ (\ref{Bfieldexp}).
    
    A straight forward analysis reveals the following set of equations at order $\gamma\kappa$
    \begin{align}
      \omega_0\Phi_2''+\omega_2\Phi_0''-\Phi_0'\omega_2'-\Phi_2'\omega_0'+4\omega_0\xi_1'\xi_2'+\frac{4}{\omega_0}\dot{\chi_1}\dot{\chi_2}-f_0=&\hspace{1mm}0,\label{twoeqi}\\
       \omega_2''-\omega_2\lambda^2-\frac{2}{\omega_0}A_{t1}'A_{t2}'+\frac{s_0 A_{t1}'^2}{\omega_0\sqrt{s_0^2+p_0^2}}=&\hspace{1mm}0,\\
       \partial_{\eta}\Bigg[\frac{\Phi_0}{\omega_0}\Bigg(A_{t2}'-\frac{s_0}{\sqrt{s_0^2+p_0^2}}A_{t1}'\Bigg)\Bigg]=&\hspace{
      1mm}0,\\
      \partial_{\eta}\Bigg[\omega_0\xi_2'-\frac{\Big(s_0\xi_1'\omega_0-p_0\dot{\chi_1}\Big)}{\sqrt{s_0^2+p_0^2}}\Bigg]=&\hspace{1mm}0,\\
      \ddot{\chi_2}=&\hspace{1mm}0,\label{twoeqf}
    \end{align}
    where we identify the above functions as
    \begin{align}
        s_0=&\hspace{1mm}-\frac{1}{\omega_0^2}A_{t1}'^2+\frac{1}{\Phi_0}\Bigg(-\frac{\dot{\chi_1^2}}{\omega_0^2}+\xi_1'^2\Bigg)\hspace{1mm},\hspace{2mm}p_0=-\frac{2}{\Phi_0\omega_0}\dot{\chi_1}\xi_1',\\
       f_0=&\hspace{1mm} \frac{2\omega_0s_0}{\sqrt{s_0^2+p_0^2}}\Bigg(\xi_1'^2+\frac{\dot{\chi_1}^2}{\omega_0^2}\Bigg).
    \end{align}
    
   The above set of equations (\ref{twoeqi})-(\ref{twoeqf}) are difficult to solve for generic values of $\eta$. However, for our present purpose, it will be sufficient to solve them near the asymptotic limit ($\eta\rightarrow\infty$) of the space-time which yields 
    
    \begin{align}
   \Phi_2=&\hspace{1mm}\frac{1}{\lambda }\Big(b_2 e^{\eta  \lambda }-b_3 \lambda +e^{-\eta  \lambda }\Big)-\frac{b_1 }{a_1}\eta  e^{-\eta  \lambda },\label{soltwoi}\\
   \omega_2=&\hspace{1mm}e^{-\eta  \lambda } \left(a_4 e^{2 \eta  \lambda }+a_5+\eta  \lambda \right),\\
    \xi_2=&\hspace{1mm}\frac{e_3 }{\lambda }e^{-\eta  \lambda }+e_4,\\
    A_{t2}=&\hspace{1mm}c_3\eta\lambda+c_4,\\
     \chi_2=&\hspace{1mm}d_3t+d_4.\label{soltwof}
    \end{align}
where $a_i,b_j,c_k,d_k$ and $e_k$, $(i=4,5,\hspace{1mm}j=2,3,\hspace{1mm}k=3,4)$ are the integration constants.

As we show below, not all of these integration constants are actually important for our analysis. In fact, a few of them finally survive which can be fixed by making use of the residual gauge freedom \cite{Castro:2008ms} in the Fefferman-Graham gauge (\ref{fggauge}). In particular,  the re-scaling of the time coordinate $t\rightarrow a_1t$ preserves the gauge condition $g_{\eta t}=0$ and $g_{\eta\eta}=1$. Therefore, we can use this freedom to fix the constant\footnote{Interestingly, with this particular choice of the integration constant $a_1$ (\ref{constc1fix}), the final expression of the central charge (\ref{cfinalexp}) appears to be independent of all the remaining integration constants. }
\begin{align}\label{constc1fix}
    a_1=\frac{1}{a_2b_3}.
\end{align}

\subsection{Order $\kappa^2$  solution }\label{quadeomap}
Finally, we estimate the quadratic order ($\kappa^2$) corrections due to the Maxwell (\ref{ONLYMAXWELLACT}) term alone.

The resulting equations of motion (\ref{geneomi})-(\ref{geneomf}) can be expressed as
\begin{align}
    \omega_0\partial_{\eta}\Bigg(\frac{\Phi_0A_{t3}'}{\omega_0}-\frac{\Phi_0\omega_1A_{t1}'}{\omega_0^2}+\frac{\Phi_1A_{t1}'}{\omega_0}\Bigg)-\omega_1\partial_{\eta}\Bigg(\frac{\Phi_0A_{t1}'}{\omega_0}\Bigg)=&\hspace{1mm}0,\label{quadkappaapi}\\
    \omega_0\Phi_3''+\omega_1\Phi_1''+\omega_3\Phi_0''-\big(\Phi_0'\omega_3'+\Phi_1'\omega_1'+\Phi_3'\omega_0'\big)+f_2=&\hspace{1mm}0,\\
    \omega_3''-\lambda^2\omega_3-\frac{1}{\omega_0}\Bigg(2A_{t1}'A_{t3}'-\frac{\omega_1}{\omega_0}A_{t1}'^2\Bigg)=&\hspace{1mm}0,\\
    \partial_{\eta}\big(\omega_0\xi_3'+\omega_1\xi_1'\big)-\frac{\omega_1}{\omega_0}\partial_{\eta}\big(\omega_0\xi_1'\big)=&\hspace{1mm}0,\\
    \ddot{\chi}_3=&\hspace{1mm}0,\label{quadkappaapf}
\end{align}
where
\begin{align}
    f_2=2\big(2\omega_0\xi_1'\xi_3'+\omega_1\xi_1'^2\big)+
\frac{2}{\omega_0}\Bigg(2\dot{\chi}_1\dot{\chi}_3-\dot{\chi}_1^2\frac{\omega_1}{\omega_0}\Bigg).
\end{align}

The above set of equations (\ref{quadkappaapi})-(\ref{quadkappaapf}) could be solved near the asymptotics ($\eta\rightarrow\infty$) of the spacetime which yield
\begin{align}
    \Phi_3= &\hspace{1mm}-\frac{1}{2
   a_1^2 \lambda ^2}\Big(e^{-\eta  \lambda } (2 \eta  \lambda +3) \left(2 a_2 a_3 \lambda -b_1 c_1 c_5\right)\Big)+\frac{b_4 e^{\eta  \lambda }}{\lambda }+\frac{a_1 \left(a_1^2+1\right)+a_3}{6 a_1^2 \left(a_1^2+1\right) a_2},\\
   \omega_3=&\hspace{1mm}\frac{1}{4} e^{-\eta  \lambda } \left(\frac{c_1 (2 \eta  \lambda +1) \left(a_3 c_1 \lambda -2
   a_1 c_5\right)}{a_1^2 \lambda }+4 a_6 e^{2 \eta  \lambda }+4 a_7\right),\\
   \xi_3=&\hspace{1mm} \frac{1}{36 a_1^{5/2} \lambda }\Big(\sqrt{a_2} e_1 e^{-3 \eta  \lambda } \left(c_1^2 (1-6 \eta  \lambda )-12 a_2
   a_3\right)\Big)-\frac{e_5 e^{-\eta  \lambda }}{\lambda }+e_6,\\
   A_{t3}=&\hspace{1mm}\frac{c_1 e^{-\eta  \lambda }}{4 a_1 a_2 b_1 \lambda }+c_6 \eta +c_5,\\
   \chi_3=&\hspace{1mm}d_5 t+d_6,
\end{align}
where $a_i,b_j,c_k,d_k$ and $e_k$ ($i=6,7,\hspace{1mm}j=4,5,\hspace{1mm}k=5,6$) are the integration constants.
     
%%%%%%%%%%%%%%%%%%%%%%%%%%%%%%%%%%%%%%%%%%%%%%%

\section{Boundary stress tensor and central charge}\label{secstcc}
In this Section, we work out the ``renormalised'' boundary stress tensor \cite{Rathi:2021aaw}, \cite{Castro:2008ms}, \cite{Cadoni:2000gm}-\cite{Hartman:2008dq} and study its transformation properties under both the diffeomorphism and the $U(1)$ gauge transformations. In particular, we examine the effects of the projected ModMax interactions on the central charge of the boundary theory.  

To begin with, we workout the boundary terms\footnote{The boundary in the Ferrerman-Graham gauge is located near $\eta\rightarrow\infty$.} for the action (\ref{lmd2d}). This is required in order to implement a consistent variational principle \cite{Rathi:2021aaw}, \cite{Castro:2008ms}. Systematically, one can decompose the boundary terms into following two pieces,
\begin{align}
    I_{\text{boundary}}&\hspace{1mm}=I_{\text{GHY}}+I_{\text{counter}},\label{fullba}
    \end{align}
    where $I_{\text{GHY}}$ is the standard Gibbons-Hawking-York boundary term and $I_{\text{counter}}$ represents the boundary counter terms.

    The Gibbons-Hawking-York boundary term \cite{Rathi:2021aaw}, \cite{Castro:2008ms}, \cite{Gibbons:1976ue} in 2D gravity is given by
    
    \begin{align}
        I_{\text{GHY}}=\frac{1}{8\pi G_2}\int_{0}^{\beta}dt\sqrt{-h}\Phi K\hspace{1mm},\hspace{2mm}K=\frac{1}{2}h^{tt}\partial_{\eta}h_{tt},
    \end{align}
    where $K$ is the trace of extrinsic curvature, $\beta$ is the inverse temperature  and $h_{tt}$  is the induced metric on the boundary. 

    On the other hand, the counter term that is required to absorb all the near boundary divergences of the on-shell action can be expressed as 
    
    \begin{align}\label{countertermdefi}
        I_{\text{counter}}=-\frac{1}{8\pi G_2}\int_{0}^{\beta}dt\sqrt{-h}\Bigg(\lambda\Phi+2\kappa\frac{b_1}{c_1} \sqrt{-h^{ab}A_aA_b}\Bigg),
    \end{align}
    where $(a,b)$ are the one dimensional boundary indices\footnote{Here, we set the constant $c_6=-\frac{\pi c_1^3}{16\sqrt{a_1a_2^3}b_1(a_1-c_1^2)}$ in order to cancel the boundary divergences up to quadratic order ($\gamma\kappa$ and $\kappa^2$) in the couplings.}. 
    
    Finally, the complete renormalised action is given by 
    \begin{align}
        I_{\text{renormalised}}=I_{\text{bulk}}+I_{\text{boundary}},\label{renormaction}
    \end{align}
    where $I_{\text{bulk}}$ and $I_{\text{boundary}}$ are given in (\ref{lmd2d}) and (\ref{fullba}) respectively.

    Notice that, the combination of the $U(1)$ gauge field in the $I_{\text{counter}}$ (\ref{countertermdefi}) seems to break the gauge invariance under the transformation 
    \begin{align}\label{u1gaugetrans}
        A_\alpha\rightarrow A_{\alpha}+\partial_{\alpha}\Sigma,
    \end{align}
    which yields the following extra piece under the $U(1)$ gauge (\ref{u1gaugetrans}) 
    \begin{align}\label{countergaugebreak}
        I_{\text{counter}}\sim \int_{0}^{\beta}dt\sqrt{-h}\Big(\sqrt{-h^{ab}A_aA_b}\Big)\rightarrow\int_{0}^{\beta}dt(A_t+\partial_t\Sigma).
    \end{align}

    However, one can preserve the gauge invariance by imposing the condition that $\partial_t\Sigma$ (see (\ref{sigmavalue})) must vanish near the boundary, $\eta\rightarrow\infty$ \cite{ Castro:2008ms}.

    Using the renormalised action (\ref{renormaction}), it is now straightforward to calculate the variation $\delta I_{\text{boundary}}$ under the combined action of the diffeomorphism and the $U(1)$ gauge, where $\delta I_{\text{boundary}}$ can be systematically expressed as\footnote{$\delta I_{\text{boundary}}$ already incorporates the bulk contributions $(\delta I_{\text{bulk}})$ near the asymptotic limit, $\eta\rightarrow\infty$. }
    \begin{align}
        \delta I_{\text{boundary}}= \frac{1}{16\pi G_2}\int dt\sqrt{-h}\Big(\mathcal{G}^{ab}\delta h_{ab}+\mathcal{G}_{\Phi}\delta \Phi+\mathcal{G}^a\delta A_a+\mathcal{G}_{\chi}\delta\chi+\mathcal{G}_{\xi}\delta\xi\Big).
    \end{align}
    
    Here, the boundary contributions can be expressed as
    \begin{align}
        \mathcal{G}^{ab}=&\hspace{1mm}n_{\mu}\nabla^{\mu}\Phi h^{ab}+n^{\mu}\frac{\Phi}{\sqrt{-h}}\Big(\partial_{\mu}\sqrt{-h}\Big)h^{ab}-\lambda\Phi h^{ab}-2\kappa \frac{b_1}{c_1} h^{ab}\sqrt{-h^{cd}A_cA_d}\nonumber\\
        &+2\kappa \frac{b_1}{c_1}\frac{A^aA^b}{\sqrt{-h^{cd}A_cA_d}},\label{stresstensordef}\\
        \mathcal{G}^a=&\hspace{1mm}-4\kappa n_{\mu}\Phi\Bigg(\cosh{\gamma}-\frac{s}{\sqrt{s^2+p^2}}\sinh{\gamma}\Bigg)F^{\mu a}+4\kappa \frac{b_1}{c_1} \frac{h^{ab}A_b}{\sqrt{-h^{cd}A_cA_d}},\\
         \mathcal{G}_{\chi}=&\hspace{1mm}-4\kappa n_{\mu}\Bigg(\nabla^{\mu}\chi\cosh{\gamma}-\frac{s\nabla^{\mu}\chi-p\epsilon^{\mu a}\nabla_{a}\xi}{\sqrt{s^2+p^2}}\sinh{\gamma}\Bigg),\\
        \mathcal{G}_{\xi}=&\hspace{1mm}-4\kappa n_{\mu}\Bigg(\nabla^{\mu}\xi\cosh{\gamma}-\frac{s\nabla^{\mu}\xi+p\epsilon^{\mu a}\nabla_{a}\chi}{\sqrt{s^2+p^2}}\sinh{\gamma}\Bigg),\\
         \mathcal{G}_{\Phi}=&\hspace{1mm}2K-2\lambda,
    \end{align}
     where $n^{\mu}=\delta^{\mu}_{\eta}$ is the unit normal vector at the boundary.

With all these preliminaries, we now introduce the boundary stress tensor \cite{Rathi:2021aaw}, \cite{Castro:2008ms} corresponding to the action (\ref{renormaction})
\begin{align}\label{rstressdef}
    T^{ab}=\frac{2}{\sqrt{-h}}\frac{\delta I_{\text{boundary}}}{\delta h_{ab}}=\frac{\mathcal{G}^{ab}}{8\pi G_2},
\end{align}
where $\mathcal{G}^{ab}$ is given in (\ref{stresstensordef}).
     
Our next task is to explore the transformation properties of the background fields (\ref{Afieldexp})-(\ref{Bfieldexp}) and hence the boundary stress tensor (\ref{rstressdef}) under the combined effects of the diffeomorphism and the $U(1)$ gauge transformation.

Under the diffeomorphism,
\begin{align}\label{ctdef}
    x^{\mu}\rightarrow x^{\mu}+\epsilon^{\mu} (x),
\end{align}
the background fields  (\ref{Afieldexp})-(\ref{Bfieldexp}) transform as 
\begin{align}
    \delta_{\epsilon}A_{\mu}=&\hspace{1mm}\epsilon^{\nu}\nabla_{\nu}A_{\mu}+A_{\nu}\nabla_{\mu}\epsilon^{\nu},\label{diffeotranscoordi}\\
\delta_{\epsilon}g_{\mu\nu}=&\hspace{1mm}\nabla_{\mu}\epsilon_{\nu}+\nabla_{\nu}\epsilon_{\mu},\label{diffeotranscoordm}\\
\delta_{\epsilon}\mathcal{S}=&\hspace{1mm}\epsilon^{\mu}\nabla_{\mu}\mathcal{S},\label{diffeotranscoordf} 
\end{align}
where $\mathcal{S}$ collectively denotes the scalar fields $\Phi,\hspace{1mm}\xi$ and $\chi$.

The diffeomorphism parameter, $\epsilon^{\mu}(x)$ can be obtained using (\ref{diffeotranscoordm}) and the space-time metric (\ref{fggauge}), which yields the following
\begin{align}\label{parametersdiffeo}
   \epsilon_t=e^{2\eta\lambda}f(t)+\frac{1}{2\lambda}\Bigg(\frac{2 \left(a_1^2+1\right)}{3 a_1^2 \lambda }-\frac{4 a_3 \kappa
   }{3 a_1^3 \lambda }\Bigg)\partial^2_tf(t) \hspace{1mm},\hspace{2mm}\epsilon_{\eta}=\Bigg(\frac{2 \left(a_1^2+1\right)}{3 a_1^2 \lambda }-\frac{4 a_3 \kappa
   }{3 a_1^3 \lambda }\Bigg)\partial_tf(t),
\end{align}
where $f(t)$ is some function\footnote{In the Fefferman-Graham gauge \cite{Rathi:2021aaw}, \cite{fefferman}, the variation of the space-time metric (under diffeomorphism (\ref{diffeotranscoordm})) yields a set of coupled differential equations that contain the  derivatives of the diffeomorphism parameters $\epsilon_t$ and $\epsilon_{\eta}$ with respect to the variable ``$\eta$''. Therefore, the function $f(t)$ in these equations appears as an integration constant. However, one can further compute the function $f(t)$ using suitable boundary conditions for the background fields $\mathcal{A}$ (\ref{Afieldexp}) and $\mathcal{B}$ (\ref{Bfieldexp}).} of time \cite{ Castro:2008ms}.

It should be noted that, we perform all the analysis in a gauge in which one of the components of the $U(1)$ gauge field, $A_{\eta}$ is set to be zero (\ref{fggauge}). On the other hand, under the diffeomorphism (\ref{ctdef}), $A_{\eta}$ transforms as 
\begin{align}\label{sigmagaugeeqn}
    \delta_{\epsilon}A_{\eta}=A_t\partial_{\eta}\Bigg(\frac{\epsilon_t}{h_{tt}}\Bigg)\neq 0,
\end{align}
which breaks the gauge condition $A_{\eta}=0$.

In order to restore this gauge condition, we employ the $U(1)$ gauge transformation, $A_\alpha\rightarrow A_{\alpha}+\partial_{\alpha}\Sigma$ and compute the $U(1)$ gauge parameter $\Sigma$ such that $(\delta_{\epsilon}+\delta_{\Sigma})A_{\eta}=0$, which yields the following 
\begin{align}\label{sigmaintermedstep}
   \Sigma= \hspace{1mm}-\int d\eta A_t\partial_{\eta}\Bigg(\frac{\epsilon_t}{h_{tt}}\Bigg),
\end{align}
where we have used the variation (\ref{sigmagaugeeqn}).

Now, one can perform the above integration (\ref{sigmaintermedstep}) using the background fields (\ref{Afieldexp})-(\ref{Bfieldexp}) and the diffeomorphism parameter (\ref{parametersdiffeo}), which yields
\begin{align}\label{sigmavalue}
    \Sigma=&\hspace{1mm}\frac{e^{-2 \eta  \lambda } }{12 a_1^5 a_2 \lambda ^3}\Bigg(f''(t) \big(2
   \left(a_1^2+1\right) a_2 a_1 \big(c_1 \lambda  \left(2 \log
   \left(a_1\right)-(\gamma -1) (2 \eta  \lambda +1)\right)+2 \gamma
    c_3 \lambda +\nonumber\\
    &\hspace{1mm}\kappa  \left(2 \lambda  \left(c_5 \eta
   +c_4\right)+c_5\right)\big)+a_1^2 c_1 \kappa  \lambda 
   \left(c_1^2 \log \left(a_1\right)-4 a_2 a_3\right) \left(2 \log
   \left(a_1\right)+2 \eta  \lambda +1\right)+\nonumber\\
   &\hspace{1mm}c_1 \kappa  \lambda 
   \left(c_1^2 \log \left(a_1\right)-8 a_2 a_3\right) \left(2 \log
   \left(a_1\right)+2 \eta  \lambda +1\right)\big)-3 a_1 a_2 c_1^3
   \kappa  \lambda ^3 f(t) \big(2 \log \left(a_1\right)+\nonumber\\
   &\hspace{1mm}2 \eta 
   \lambda +1\big)\Bigg).
\end{align}
It is interesting to notice that the $U(1)$ gauge parameter $\Sigma$ vanishes naturally in the asymptotic limit ($\eta\rightarrow\infty$), which is consistent with the gauge preserving condition (\ref{countergaugebreak}).

Finally, we note down the transformation of the boundary stress tensor (\ref{rstressdef}) under the combined action of the diffeomorphism (\ref{ctdef}) and the $U(1)$ gauge transformation which yields
\begin{align}
    (\delta_{\epsilon}+\delta_{\Sigma})T_{tt}=&\hspace{1mm}\frac{1}{8\pi G_2}\Bigg[\Bigg(\partial_{\eta}\Phi-\lambda\Phi-2\kappa\frac{b_1}{c_1}\frac{ A_t}{\omega}\Bigg)(\delta_{\epsilon} h_{tt})+4\kappa\frac{b_1\omega}{c_1} \big((\delta_{\epsilon}+\delta_{\Sigma}) A_t\big)\nonumber\\
    &+\frac{\Phi}{2}\partial_{\eta}(\delta_{\epsilon} h_{tt})-\frac{1}{2}\Big(\partial_{\eta}\omega^2-2\lambda\omega^2\Big)(\delta_{\epsilon}\Phi)-\partial_{\eta}(\delta_{\epsilon}\Phi)\omega^2\Bigg].\label{variationstresstensor}
\end{align}

 The variations of the background fields $h_{tt}, \hspace{1mm} A_t$ and $ \Phi$ can be obtained using (\ref{diffeotranscoordi})-(\ref{parametersdiffeo}) and (\ref{sigmavalue}), which yields the following

\begin{align}
(\delta_{\epsilon}+\delta_{\Sigma})A_t=&\hspace{1mm}\mathcal{H}_1(\eta)\partial_t f(t)+\mathcal{H}_2(\eta)\partial_t^3 f(t)\label{fv1},\\
   \delta_{\epsilon} h_{tt} =&\hspace{1mm} \mathcal{H}_3(\eta) \partial_tf(t) +\mathcal{H}_4(\eta)\partial_t^3f(t),\\
   \delta_{\epsilon} \Phi =&\hspace{1mm}\mathcal{H}_5(\eta)\partial_t f(t)\label{fv2},
\end{align}
where the explicit form of the functions $\mathcal{H}_i(\eta)$, $(i=1,2...5)$ are given in the Appendix \ref{h1h2fucdetailap}.

Using these variations (\ref{fv1})-(\ref{fv2}), the transformation of the boundary stress tensor (\ref{variationstresstensor}) can be expressed in a more elegant way 
\begin{align}
    (\delta_{\epsilon}+\delta_{\Sigma})\Tilde{T}_{tt}=2\Tilde{T}_{tt}\partial_tf(t)+f(t)\partial_t\Tilde{T}_{tt}- c_M\partial^3_tf(t).\label{stvariation}
\end{align}

Here, we define the re-scaled stress tensor as 
\begin{align}
    \Tilde{T}_{tt}= \frac{T_{tt}}{b_3(1+a_2^2b_3^2)},
\end{align}
 and identify the coefficient ``$c_M$'' (coefficient of $\partial_t^3f(t)$) as being the central charge \cite{Rathi:2021aaw}, \cite{Castro:2008ms} of the boundary theory,
\begin{align}\label{cfinalexp}
    c_M=\frac{1}{144 \sqrt{3} \pi  G_2}\Big(\kappa-12 \gamma \kappa +2 \kappa^2\Big),
\end{align}
where we substitute $\lambda=\sqrt{3}$.

It should be noted that the above expression of the central charge (\ref{cfinalexp}) is a perturbative result up to quadratic order in the ModMax coupling ($\gamma$) and the $U(1)$ gauge coupling ($\kappa$). Clearly, in the limit $\gamma\rightarrow 0$, the central charge (\ref{cfinalexp}) reduces to $\sim\frac{1}{G_2}$ which is consistent with the existing result in the literature \cite{Castro:2008ms}.
%%%%%%%%%%%%%%%%%%%%%%%%%%%%%%%%%%%%
\section{Black holes and 2D projected ModMax  }\label{secbhwithmm}
We now construct the 2D black hole solutions and investigate their thermal properties in the presence of 2D projected ModMax interactions (\ref{lmd2d}). In particular, we emphasise on the role played by the ModMax parameter, that is required to set all the fields ``finite'' near the horizon. These solutions are further used to compute the Wald entropy \cite{Wald:1993nt}-\cite{Pedraza:2021cvx} associated with these 2D black holes. Finally, we also comment on the possibilities for extremal black hole solutions in two dimensions. 

\subsection{Black hole solutions}
We estimate the 2D black hole solutions of (\ref{lmd2d}) by means of  perturbative techniques up to quadratic order in the ModMax parameter $(\gamma)$ and the Maxwell's coupling ($\kappa$). Technically speaking, it is not convenient to determine the black hole horizon in the Ferrferman-Graham gauge due to the presence of the non-trivial couplings in $U(1)$ gauge fields (\ref{lmd2d}). However, one can perform an elegant calculation using the light cone gauge. In this gauge, the space-time metric can be expressed as  
\begin{align}\label{bhlcgauge}
    ds^2=&\hspace{1mm}e^{2\omega(z)}\big(-dt^2+dz^2\big),\hspace{1mm}A_{\mu}dx^{\mu}=\hspace{1mm}A_t(z)dt,\nonumber\\
       \Phi=\hspace{1mm}&\Phi(z),\hspace{1mm}\chi=\chi(t),\hspace{1mm}\xi=\xi(z).
\end{align}

Like before as in (\ref{Afieldexp})-(\ref{Bfieldexp}), one can systematically expand the background fields in the couplings $\kappa$ and $\gamma$ as
\begin{align}
   & \mathcal{A}^{(bh)}=\mathcal{A}_0^{(bh)}+\kappa\mathcal{A}_1^{(bh)}+\gamma\kappa\mathcal{A}_2^{(bh)}+\kappa^2\mathcal{A}_3^{(bh)}...,\label{fieldexpansionblackhole1}\\
   & \mathcal{B}^{(bh)}=\mathcal{B}_1^{(bh)}+\gamma\mathcal{B}_2^{(bh)}+\kappa \mathcal{B}_3^{(bh)}...,\hspace{2mm}|\kappa|<<1,\hspace{1.5mm}|\gamma|<<1,\label{fieldexpansionblackhole2}
\end{align}
where $\mathcal{A}^{(bh)}$ collectively represents the fields ($\Phi$, $\omega$) and $\mathcal{B}^{(bh)}$ represents the remaining fields ($A_t$, $\chi$, $\xi$). Furthermore, the superscript ``${bh}$'' in $\mathcal{A}^{(bh)}$ and $\mathcal{B}^{(bh)}$ denote the black hole solution.
\subsubsection{Zeroth order solution}
In order to calculate black hole solutions at zeroth order, we switch off the $U(1)$ gauge couplings ($\kappa\rightarrow0$, $\gamma\rightarrow0$) in the equations of motion (\ref{geneomi})-(\ref{geneomf}), which yields the following set of equations
\begin{align}
    \Phi_0''-\omega_0'\Phi_0'+\Lambda e^{2\omega_0}\Phi_0=&\hspace{1mm}0,\label{bhzeroeomi}\\
    \omega_0'\Phi_0'+\Lambda e^{2\omega_0}\Phi_0=&\hspace{1mm}0,\\
     \omega_0''+e^{2\omega_0}\Lambda=&\hspace{1mm}0,\label{bhzeroeomf}
\end{align}
where $'$ denotes the derivative with respect to $z$.

On solving the equations (\ref{bhzeroeomi})-(\ref{bhzeroeomf}), one finds 
\begin{align}\label{bhzerosollc}
    e^{2\omega_{0}^{(bh)}}=-\frac{4\mu}{\Lambda \sinh^2{(2\sqrt{\mu}}z)}\hspace{1mm},\hspace{2mm}\Phi_0^{(bh)}=\phi_0,
\end{align}
where $\phi_0$ is a constant. 

It should be noted that we treat the dilaton ($\Phi$) as constant while taking the limits $\kappa\rightarrow0$ and $\gamma\rightarrow0$. However, it possesses a non-trivial profile in the presence of $U(1)$ gauge fields (see Section (\ref{subsecbhk}) and (\ref{subsecbhg})).

\subsubsection{Order $\kappa$ solution}\label{subsecbhk}
The leading order corrections to $\mathcal{A}^{(bh)}$ and $\mathcal{B}^{(bh)}$ could be estimated by solving the equations of motion (\ref{geneomi})-(\ref{geneomf}) at order $\kappa$

\begin{align}
\Phi_1''-2\big(\omega_0'\Phi_1'+\omega_1'\Phi_0'\big)+2\big(\dot{\chi}_1^2+\xi_1'^2\big)=&\hspace{1mm}0,\label{bhoneeomi}\\
\omega_1''+2\Lambda\omega_1e^{2\omega_0}-e^{-2\omega_0}A_{t1}'^2=&\hspace{1mm}0,\\
    \partial_{z}\Big(\Phi_0e^{-2\omega_0}A_{t1}'\Big)=&\hspace{1mm}0,\\
    \xi_1''=&\hspace{1mm}0,\\
    \ddot{\chi}_1=&\hspace{1mm}0,\label{bhoneeomf}
\end{align}
where . and $'$ denote the derivatives with respect to $t$ and $z$ respectively. 

In order to solve the above differential equations (\ref{bhoneeomi})-(\ref{bhoneeomf}), we adopt the following change in coordinates 
\begin{align}\label{bhctrhoz}
    \rho=\sqrt{\mu}\coth{(2\sqrt{\mu}z)}.
\end{align}

Using the zeroth order solutions (\ref{bhzerosollc}) together with (\ref{bhctrhoz}), one finds
\begin{align}
    \omega_1^{(bh)}=&\hspace{1mm}\frac{q_2}{\sqrt{\mu}}\rho\tanh^{-1}{\Bigg(\frac{\rho}{\sqrt{\mu}}\Bigg)}+\frac{q_1\rho}{\sqrt{\mu}}+\frac{m_1^2}{2\Lambda\phi_0^2}-q_2,\label{bhsolonei}\\
    \Phi_1^{(bh)}=&\hspace{1mm}-\frac{(n_1^2+l_1^2)\rho}{4\mu^{\frac{3}{2}}}\tanh^{-1}{\Bigg(\frac{\rho}{\sqrt{\mu}}\Bigg)}+g_1\rho,\label{phi1bh}\\
 \xi_1^{(bh)}=&\hspace{1mm}\frac{l_1}{2\sqrt{\mu}}\coth^{-1}
    {\Bigg(\frac{\rho}{\sqrt{\mu}}\Bigg)}+l_2\label{xi1bh},\\
      A_{t1}^{(bh)}=&\hspace{1mm}\frac{2m_1\rho}{\Lambda\phi_0}+m_2,\\
    \chi_1^{(bh)}=&\hspace{1mm}n_1t+n_2\label{bhsolonef},
\end{align}
where $m_i,n_i,l_i,q_i$ and $g_1$, $(i=1,2)$ are the integration constants.

\subsubsection{Order $\gamma\kappa$  solution}\label{subsecbhg}
The contributions due to the projected ModMax interactions could be estimated by solving the equations of motion (\ref{geneomi})-(\ref{geneomf}) at order $\gamma\kappa$

\begin{align}
\Phi_2''-2\big(\omega_0'\Phi_2'+\omega_2'\Phi_0'\big)+4\big(\dot{\chi}_1\dot{\chi}_2+\xi_1'\xi_2'\big)-\frac{2s_0}{\sqrt{s_0^2+p_0^2}}\big(\dot{\chi}_1^2+\xi_1'^2\big)=&\hspace{1mm}0,\label{bhtwoeomi}\\
\omega_2''+2\Lambda\omega_2e^{2\omega_0}-2e^{-2\omega_0}A_{t1}'A_{t2}'-e^{2\omega_0}\sqrt{s_0^2+p_0^2}+f_1=&\hspace{1mm}0,\\
    \partial_z\Bigg[\xi_2'-\frac{1}{\sqrt{s_0^2+p_0^2}}\Big(s_0\xi_1'-p_0\dot{\chi}_1\Big)\Bigg]=&\hspace{1mm}0,\\
    \partial_z\Bigg[e^{-2\omega_0}\Phi_0\Bigg(A_{t2}'-\frac{A_{t1}'s_0}{\sqrt{s_0^2+p_0^2}}\Bigg)\Bigg]=&\hspace{1mm}0,\\
    \ddot{\chi}_2=&\hspace{1mm}0,\label{bhtwoeomf}
\end{align}
where we define the above quantities as 
\begin{align}
   f_1=&\hspace{1mm}\frac{1}{\Phi_0\sqrt{s_0^2+p_0^2}}\Big(s_0\big(-\dot{\chi}_1^2+\xi_1'^2\big) -2p_0\dot{\chi}_1\xi_1'\Big)\hspace{1mm},\hspace{2mm} p_0=-2\Phi_0^{-1}e^{-2\omega_0}\dot{\chi}_1\xi_1',\nonumber\\ s_0=&\hspace{1mm}-e^{-4\omega_0}A_{t1}'^2+\Phi_0^{-1}e^{-2\omega_0}\big(-\dot{\chi}_1^2+\xi_1'^2\big).  
\end{align}

Clearly, the above differential equations (\ref{bhtwoeomi})-(\ref{bhtwoeomf}) are quite non trivial to solve exactly in the radial variable ($z$). However, for the purpose of our present analysis, it is sufficient to solve them near the black hole horizon.

 Using (\ref{bhctrhoz}), the location of the horizon $(\rho_H)$ can be determined by noting the spacetime metric (\ref{bhlcgauge})
\begin{align}\label{expbhmetric}
    ds^2_{(bh)}\approx\frac{4(\mu-\rho^2)}{\Lambda}\Big(1+2\kappa\omega_1^{(bh)}+2\gamma\kappa\omega_2^{(bh)}\Big)\Bigg(-dt^2+\frac{d\rho^2}{4(\mu-\rho^2)^2}\Bigg),
\end{align}
 which yields $\rho=\rho_H=\sqrt{\mu}$.

Finally, the near horizon solutions of the equations of motion (\ref{bhtwoeomi})-(\ref{bhtwoeomf}) could be listed as
\begin{align}
\Phi_2^{(bh)}=&\hspace{1mm}\frac{\rho  }{192 \mu ^{5/2}}\Bigg(64 \mu  \rho  \left(\rho -6 \sqrt{\mu }\right)+n_1^2 \rho  \Bigg(\frac{64 \mu ^{3/2} }{n_1^8l_1^2} \Bigg(8 \sqrt{\mu } (n_1^2-l_1^2)^2 \left(n_1^2+l_1^2\right) \left(\rho -9 \sqrt{\mu
   }\right)\nonumber\\
   &\hspace{1mm}-3n_1^4l_1^4\Bigg)-15 \sqrt{\mu
   }+2
   \rho \Bigg)+2 n_1 n_3 \rho  \left(2 \rho -15 \sqrt{\mu }\right)+\rho  l_1 \Big(l_1
   \left(\rho -12 \sqrt{\mu }\right)+\nonumber\\
   &3 l_3 \left(\rho -9 \sqrt{\mu }\right)\Big)+192 \mu
   ^{5/2} g_2\Bigg)+\frac{\rho  \left(n_1^2+2 n_3 n_1+l_1 \left(l_1+2
   l_3\right)\right) \log \left(\rho -\sqrt{\mu }\right)}{8 \mu ^{3/2}}\label{phi2bh},\\
   \xi_2^{(bh)}=&\hspace{1mm}\frac{1}{16 \mu ^{3/2}}\Bigg(-\frac{256 m_1^4 n_1^2 l_1 \mu ^2 \rho  \left(\rho -2 \sqrt{\mu }\right)}{\Lambda ^2 \text{$\phi_0 $}^2(n_1^2+l_1^2)^3}-\frac{32 \mu  \rho  \left(2 \sqrt{\mu }
   \left(n_1^2-2 l_1^2\right)+l_1^2 \rho \right)}{n_1^2l_1}+\nonumber\\
    &\hspace{1mm}(l_{3}+l_1)\Bigg(-\frac{1}{2} \rho\left(\rho -6
   \sqrt{\mu }\right)-4 \mu \log \left(\rho -\sqrt{\mu }\right)\Bigg)\Bigg)+l_{4},\label{xi2bh}\\
    A_{t2}^{(bh)}=&\hspace{1mm}\frac{32 \sqrt{\mu } m_1^3 n_1^2  l_1^2  \rho\left(\rho -2 \sqrt{\mu }\right)}{\Lambda ^2 \text{$\phi_0$}^2(n_1^2+l_1^2)^3
   }-m_3 \rho +m_4,\\
   \omega_2^{(bh)}=&\hspace{1mm}\frac{m_1 }{2
   \text{$\phi_0$}^2}\left(\frac{m_1 \left(n_1^2-l_1^2\right)}{\Lambda (n_1^2+l_1^2) }-m_3 \text{$\phi_0 $}\right)+q_3 I_0\left(\Tilde{\rho}\right)+q_4 K_0\left(\Tilde{\rho}\right),\\
   \chi_2^{(bh)}=&\hspace{1mm}n_3t+n_4,\label{chi2bh}
\end{align}
where we define $\Tilde{\rho}=2 \sqrt{\frac{\rho }{\sqrt{\mu }}-1}$ and $m_i,n_i,l_i,q_i$, $g_2$, $(i=3,4)$ are the integration constants. Furthermore, here $I_0\left(\Tilde{\rho}\right)$ and $ K_0\left(\Tilde{\rho}\right)$ are respectively the modified Bessel functions \cite{mpfunction} of the first $(I_{n}(\Tilde{\rho}))$ and the second kind $(K_{n}(\Tilde{\rho}))$.

\subsubsection{Order $\kappa^2$  solution }
The contribution due to the Maxwell (\ref{ONLYMAXWELLACT}) term alone at quadratic ($\kappa^2$) could be estimated by solving the equations (\ref{geneomi})-(\ref{geneomf}) at order $\kappa^2$ 
\begin{align}
-2\omega_1\partial_z\Big(\Phi_0e^{-2\omega_0}A_{t1}'\Big)+\partial_z\Big[e^{-2\omega_0}\Big(-2\omega_1\Phi_0A_{t1}'+\Phi_1A{t1}'+\Phi_0A_{t3}'\Big)\Big]&=0,\label{bhjhep1eq}\\
    \omega_3''+2\Lambda e^{2\omega_0}\big(\omega_1^2+\omega_3\big)-2e^{-2\omega_0}\big(-\omega_1 A_{t1}'^2+A_{t1}'A_{t3}'\big)&=0,\\
\Phi_3''-2\big(\omega_3'\Phi_0'+\omega_0'\Phi_3'+\omega_1'\Phi_1'\big)+4(\dot{\chi}_1\dot{\chi}_3+\xi_1'\xi_3')&=0,\\
    \xi_3''&=0,\\
    \ddot{\chi}_3&=0.\label{bhjhepfeq}
\end{align}

The solutions of the above equations (\ref{bhjhep1eq})-(\ref{bhjhepfeq}) are quite complicated, therefore we mention them in the Appendix \ref{jhepappendixk2}. Like before, one can further simplify these solutions (\ref{phi2bh})-(\ref{chi2bh}) and (\ref{jhepeopmkappa2i})-(\ref{jhepeopmkappa2f}) by making use of the residual gauge freedom in the light cone gauge (\ref{expbhmetric}). In particular, the re-scaling of the time coordinate, $t\rightarrow n_1t$  does not affect the gauge condition $g_{t\rho}=0$. Therefore, one can use this freedom to fix the constant $ n_1=\sqrt{1-l_1^2}$. 

It is evident from (\ref{phi1bh}), (\ref{xi1bh}), (\ref{jhepeopmkappa2i1}) and (\ref{jhepeopmkappa2i2})  that the leading order ($\kappa$) corrections as well as the quadratic order $(\kappa^2)$ corrections diverge as we move closer towards the black hole horizon ($\rho\sim\rho_H=\sqrt{\mu}$). Similar divergences persist even at quadratic order ($\gamma\kappa$) (see (\ref{phi2bh}) and (\ref{xi2bh})). However, for a particular choice of constants 
\begin{align}\label{n3l3}
 n_3=&\hspace{1mm}\frac{1 }{2\gamma\sqrt{1-l_1^2}}\Big((2l_1^2-1)(1+\gamma)-2\kappa n_5 \sqrt{1-l_1^2}-\kappa q_2\Big),\\
 l_3=&\hspace{1mm}\frac{1}{\gamma l_1}\Big(-l_1^2(1+\gamma)+\kappa n_5\sqrt{1-l_1^2}\Big),
\end{align}
the divergences at order $\kappa$ and $\kappa^2$ cancel with those at the quadratic order $(\gamma \kappa)$ thereby resulting in a finite expression for $\xi^{(bh)}$ and $\Phi^{(bh)}$ near the horizon $(\rho\sim\sqrt{\mu})$. This turns out to be a unique feature of projected ModMax interactions in two dimensions.

\subsection{2D Black hole thermodynamics}
With the above solutions at hand, we now explore the thermal properties of 2D black holes in the presence of projected ModMax interactions. In particular, we compute the Wald entropy \cite{Wald:1993nt}-\cite{Pedraza:2021cvx} for 2D black holes. Finally, we also comment on the Wald entropy associated with the \emph{extremal} black holes in two dimensions. 

To begin with, we compute the Hawking temperature \cite{Hawking:1975vcx} for the 2D black holes which receives quadratic order corrections due to $U(1)$ gauge and ModMax couplings
\begin{align}\label{hawktdef}
    T_H=\frac{1}{2\pi}\sqrt{-\frac{1}{4}g^{tt}g^{\rho\rho}\big(\partial_{\rho}g_{tt}\big)^2}\Bigg|_{\rho\rightarrow\sqrt{\mu}}=\frac{\sqrt{\mu}}{\pi}\Bigg[1-\Big(\kappa q+ \gamma\kappa q+\kappa^2p\Big)\Bigg],
\end{align}
where we set the constants $q_4=q_{2}=q$ and $p$ is defined as
\begin{align}\label{jpcostbh}
    p=\frac{m_1^2}{12\Lambda\mu\phi_0^3}\Big(1+24q\mu\phi_0+\log(4)-8\mu^{\frac{3}{2}}g_1\Big).
\end{align}

The Wald entropy \cite{Wald:1993nt}-\cite{Pedraza:2021cvx} is defined as
\begin{align}\label{walddefblackhole}
    S_W=-2\pi\frac{\delta \mathcal{L}}{\delta R_{\mu\nu\alpha\beta}}\epsilon_{\mu\nu}\epsilon_{\alpha\beta},
\end{align}
where $R_{\m\nu\alpha\beta}$ is the Riemann curvature tensor, $\mathcal{L}$ is the Lagrangian density\footnote{Here we used the convention, $I=\int d^2x\sqrt{-g}\mathcal{L}$. } in two dimensions and  $\epsilon_{\mu\nu}$ is the anti-symmetric rank two tensor having the normalization condition, $\epsilon^{\mu\nu}\epsilon_{\mu\nu}=-2$.  

Using (\ref{renormaction}), the Wald entropy (\ref{walddefblackhole}) for 2D black holes turns out to be\footnote{Here, the entities $\phi_1$, $\phi_2$ and $\phi_3$ are respectively the values of  $\Phi_1^{(bh)}$ (\ref{phi1bh}), $\Phi_2^{(bh)}$ (\ref{phi2bh}) and $\Phi_3^{(bh)}$ (\ref{jhepeopmkappa2i1}) at the horizon $\rho=\rho_H=\sqrt{\mu}$.}
\begin{align}\label{waldentexp01}
 S_W=\frac{\Phi^{(bh)}}{4G_2}\Bigg|_{\rho\rightarrow\sqrt{\mu}}=\hspace{1mm}\frac{1}{4G_2}\Big(\phi_0+\kappa\phi_1+\gamma\kappa\phi_2+\kappa^2\phi_3\Big),
\end{align}
where we denote the above entities as
\begin{align}
    \phi_1=&\hspace{1mm}\sqrt{\mu } g_1-\frac{1}{192 \mu }\Big(12 \log (4 \mu )+2 l_1^2-13\Big),\label{phih}\\
    \phi_2=&\hspace{1mm}\frac{64 \mu  }{3 l_1^2}\frac{\left(1-2 l_1^2\right){}^2}{\left(l_1^2-1\right){}^3}+\frac{l_1^2}{l_1^2-1}+\sqrt{\mu } g_2-\frac{5}{3}\label{ph2h},\\
    \phi_3=&\hspace{1mm}\frac{1}{192 \mu }\Bigg(192 \mu ^{3/2} \left(g_1
   \left(q_1+q\right)+g_3\right)+2 \sqrt{1-l_1^2} n_5+q
   (36 \log (\mu )\nonumber\\
   &+13-24 \log (2))-48 q_1\Bigg),\label{jph3h}
\end{align}
and $\phi_0$ is the usual constant dilaton solution in the limit $\kappa\rightarrow0$ and $\gamma\rightarrow0$ (\ref{bhzerosollc}). 
 
 \subsection{A special case : Extremal 2D black holes}

As a special case, we study the extremal 2D black hole solutions and compute the associated Wald entropy. Extremal black holes correspond to the vanishing of the Hawking temperature (\ref{hawktdef}) 
\begin{align}\label{extremalconstraint}
    \kappa q+ \gamma\kappa q+\kappa^2p=1,
\end{align}
which for the present example stands as an extremality condition in two dimensions.

Using (\ref{extremalconstraint}) and (\ref{waldentexp01}), the Wald entropy for 2D extremal black holes $\big(S_{W}^{\text{(ext)}}\big)$ turns out to be  
  \begin{align}
      S_{W}^{\text{(ext)}}=\frac{1}{4G_2}\Bigg[\phi_0+\frac{\phi_2}{q}+\kappa\Big(\phi_1-\phi_2\Big)+\kappa^2\Bigg(\phi_3-\frac{p}{q}\phi_2\Bigg)\Bigg],
 \end{align}
 where the entities $p$, $\phi_1$, $\phi_2$ and $\phi_3$ are respectively given in (\ref{jpcostbh}), (\ref{phih}), (\ref{ph2h}) and (\ref{jph3h}).

 %%%%%%%%%%%%%%%%%%%%%%%%%%%%%%%%%%%%%%%
\section{Concluding remarks}\label{secconc}
To summarise, in the present paper, we construct the 2D analogue of the four dimensional ModMax electrodynamics (coupled with Einstein gravity) using the notion of dimensional reduction. We investigate the effects of projected ModMax interactions on various physical entities associated with the boundary theory in one dimension. Finally, we construct the associated 2D black hole solutions and explore their thermal properties.

 Below, we outline some of the future extensions of the present work.

$\bullet$  In the literature, there exists an alternative way to derive the thermodynamic entropy of 2D black holes by noting the asymptotic growth of the physical states of a CFT by means of the Cardy formula ($S_C$) \cite{Castro:2008ms}, \cite{ Cardy:1986ie}-\cite{Strominger:1997eq}
 \begin{align}\label{cardyformula}
     S_C= 2\pi\sqrt{\frac{c_M \Delta}{6}},
 \end{align}
 where $\Delta$ is the  eigen value of the associated Virasoro generator $L_0$. 
 
 The authors in  \cite{ Castro:2008ms} establish a 2D/3D dictionary which by virtue of the Cardy formula (\ref{cardyformula}) predicts the correct Bekenstein-Hawking entropy for 2D black holes. Therefore, it would be indeed an interesting project to uplift the 2D black hole solutions (\ref{fieldexpansionblackhole1})-(\ref{fieldexpansionblackhole2}) into three dimensions and establish a suitable 2D/3D mapping in the presence of 2D projected ModMax interactions.

$\bullet$ It would be an interesting project to add SU(2) Yang-Mills interactions and investigate their imprints on various physical observables associated with the boundary theory. In particular, the authors in \cite{Lala:2020lge} observe that the SU(2) Yang-Mills field play an important role in obtaining the Hawking-Page transition in the context of JT gravity. Therefore, one can investigate similar effects and/or possible deviations in the presence of projected ModMax interactions in two dimensions. 

$\bullet$ Finally, it would be nice to construct the 2D wormhole solutions \cite{Garcia-Garcia:2020ttf}, \cite{Rathi:2021mla} and explore their thermal stability for the ModMax corrected JT gravity models.

We would like to address some of the above issues in the near future.

%%%%%%%%%%%%%%%%%%%%%%%%%%%
\section*{Acknowledgments}
The authors are indebted to the authorities of Indian Institute of Technology, Roorkee for their unconditional support towards researches in basic sciences. DR would like to acknowledge The Royal Society, UK for financial assistance. 
%%%%%%%%%%%%%%%%%%%%%%%%%%%%%%%%%
%%%%%%%%%%%%%%%%%%%%%%%%%%%%%%%%%%%%%%%%%%%%%%%%%%%%%%%%%%%%

\appendix
\section{Equations of motion}\label{geneomap}
In this Appendix, we note down the most general form of the equations of motion (\ref{geneomi})-(\ref{geneomf}) in the Fefferman-Graham gauge (\ref{fggauge}),

 \begin{align}
    A_{t}\hspace{1mm}:\hspace{1mm}&\frac{\kappa}{\sqrt{-h_{tt}}}\partial_{\eta}\Bigg[\frac{\Phi A_t'}{\sqrt{-h_{tt}}}\Bigg(\cosh{\gamma}-\frac{s\sinh{\gamma}}{\sqrt{s^2+p^2}}\Bigg)\Bigg]=0,\label{gaugeeomiap}\\
    \chi\hspace{1mm}:\hspace{1mm}&\frac{\kappa}{\sqrt{-h_{tt}}}\partial_t\Bigg(-\frac{\dot{\chi}}{\sqrt{-h_{tt}}}\cosh{\gamma}+\frac{s\dot{\chi}+p\xi'\sqrt{-h_{tt}}}{\sqrt{-h_{tt}}\sqrt{s^2+p^2}}\sinh{\gamma}\Bigg)+\frac{\kappa}{\sqrt{-h_{tt}}}\partial_{\eta}\Bigg(\sqrt{-h_{tt}}\times\nonumber\\
    &\chi'\cosh{\gamma}-\frac{s\chi'\sqrt{-h_{tt}}+p\dot{\xi}}{\sqrt{s^2+p^2}}\sinh{\gamma}\Bigg)=0,\\
     \xi\hspace{1mm}:\hspace{1mm}&\frac{\kappa}{\sqrt{-h_{tt}}}\partial_t\Bigg(-\frac{\dot{\xi}}{\sqrt{-h_{tt}}}\cosh{\gamma}-\frac{-s\dot{\xi}+p\chi'\sqrt{-h_{tt}}}{\sqrt{-h_{tt}}\sqrt{s^2+p^2}}\sinh{\gamma}\Bigg)+\frac{\kappa}{\sqrt{-h_{tt}}}\partial_{\eta}\Bigg(\sqrt{-h_{tt}}\times\nonumber\\
    &\xi'\cosh{\gamma}-\frac{s\xi'\sqrt{-h_{tt}}-p\dot{\chi}}{\sqrt{s^2+p^2}}\sinh{\gamma}\Bigg)=0,
\end{align}

 \begin{align}
    \Phi\hspace{1mm}:\hspace{1mm}&\sqrt{-h_{tt}}(\sqrt{-h_{tt}})''+(\sqrt{-h_{tt}})^2\Lambda-\kappa A_t'^2\cosh{\gamma}+\frac{\kappa s A_t'^2}{\sqrt{s^2+p^2}}\sinh{\gamma}=0,\\
    g_{tt}\hspace{1mm}:\hspace{1mm}&\Phi''+\Lambda\Phi
+2\kappa\Phi\Bigg(\cosh{\gamma}-\frac{s\sinh{\gamma}}{\sqrt{s^2+p^2}}\Bigg)\Bigg(\Phi^{-1}(\chi'^2+\xi'^2)-\frac{s}{2}\Bigg)=0,\\
g_{\eta\eta}\hspace{1mm}:\hspace{1mm}&-\frac{1}{\sqrt{-h_{tt}}}\partial_{t}\Bigg(\frac{\dot{\Phi}}{\sqrt{-h_{tt}}}\Bigg)+\frac{(\sqrt{-h_{tt}})'\Phi'}{\sqrt{-h_{tt}}}+\Lambda\Phi-2\kappa\Phi\Bigg(\cosh{\gamma}-\frac{s\sinh{\gamma}}{\sqrt{s^2+p^2}}\Bigg)\times\nonumber\\
&\Bigg(\frac{\Phi^{-1}(\dot{\chi}^2+\dot{\xi}^2)}{(\sqrt{-h_{tt}})^2}+\frac{s}{2}\Bigg)=0,\\
g_{\eta t}\hspace{1mm}:\hspace{1mm}&-(\dot{\Phi})'+\frac{\dot{\Phi}(\sqrt{-h_{tt}})'}{\sqrt{-h_{tt}}}-2\kappa \Bigg(\cosh{\gamma}-\frac{s\sinh{\gamma}}{\sqrt{s^2+p^2}}\Bigg)(\dot{\xi}\xi'+\dot{\chi}\chi')=0,\label{gaugeeomfap}
\end{align}
along with the functions
\begin{align}
    s=&-\frac{A_t'^2}{(\sqrt{-h_{tt}})^2}+\Phi^{-1}\Bigg(-\frac{1}{(\sqrt{-h_{tt}})^2}(\dot{\chi}^2+\dot{\xi}^2)+\chi'^2+\xi'^2\Bigg),\\
    p=&-2\frac{\Phi^{-1}}{\sqrt{-h_{tt}}}\big(\dot{\chi}\xi'-\chi'\dot{\xi}\big),
\end{align}
where . and $'$ denote the derivatives with respect to $t$ and $\eta$ respectively. 

%%%%%%%%%%%%%%%%%%%%%%%

%%%%%%%%%%%%%%%%%%%%%%%%%%%%%%%%%%%%%%%%%
\section{Details of the functions $\mathcal{H}_i$'s}\label{h1h2fucdetailap}
In this Appendix, we present the explicit details of the functions $\mathcal{H}_i$, $(i=1,2,..5)$
\begin{align}
    \mathcal{H}_1(\eta)=&\hspace{1mm}\frac{1}{4} \Bigg[c_1 \Bigg(-\frac{a \kappa  e^{-\eta  \lambda }}{a_1 a_2 b_1}+\frac{\kappa 
   }{a_1^3}\left(a_3 \left(8 \log \left(a_2-a_1 e^{2 \eta  \lambda }\right)-8 \eta  \lambda
   \right)-\frac{e^{-\eta  \lambda }}{a_2 b_1 \lambda }\right)+\nonumber\\
   &\hspace{1mm}\frac{1}{a_1^2}\Big(4 (\gamma +1) \eta 
   \lambda -4 \log \left(a_2-a_1 e^{2 \eta  \lambda }\right)\Big)-4 a (\gamma +1) \lambda
   +\frac{8 a a_1 \lambda  e^{2 \eta  \lambda }}{a_1 e^{2 \eta  \lambda }-a_2}\Bigg)-\nonumber\\
   &\hspace{1mm}\frac{4
   }{a_1^2}\left(\kappa  \left(c_5 \left(\eta -a a_1^2\right)+c_4\right)+\gamma 
   c_3\right)-\frac{1}{a_1^4 a_2}\Big(c_1^3 \kappa  e^{-2 \eta  \lambda } \Big(2 a_1 e^{2 \eta  \lambda }\Big(2 \eta ^2 \lambda ^2+
   \nonumber\\
   &\hspace{1mm}\log \left(a_2-a_1 e^{2 \eta  \lambda }\right)^2-3 \eta  \lambda  \log \left(a_2-a_1
   e^{2 \eta  \lambda }\right)\Big)+a_2 \left(2 \log \left(a_1\right)+2
   \eta  \lambda +1\right)\Big)\Big)\Bigg],
\end{align}
\begin{align}
 \mathcal{H}_2(\eta)=&\hspace{1mm}-\frac{1}{8 a_1^3
   a_2 b_1 \lambda ^2}\Bigg(a e^{-3 \eta  \lambda } \Big(c_1 \Big(-2 a_1 a_2 b_1 \lambda  e^{\eta  \lambda }
   \Big(-2 \log \left(a_2-a_1 e^{2 \eta  \lambda }\right)+2 \log \left(a_1\right)\nonumber\\
   &\hspace{1mm}-\gamma +4
   \eta  \lambda +1\Big)+4 a_2 a_3 b_1 \kappa  \lambda  e^{\eta  \lambda } \Big(-2 \log
   \left(a_2-a_1 e^{2 \eta  \lambda }\right)+2 \log \left(a_1\right)+4 \eta  \lambda\nonumber\\
   &\hspace{1mm}
   +1\Big)+\kappa \Big)+b_1 c_1^3 \kappa  \lambda  e^{\eta  \lambda } \Big(2 \Big(\log
   ^2\left(a_2-a_1 e^{2 \eta  \lambda }\right)-3 \eta  \lambda  \log \left(a_2-a_1 e^{2 \eta 
   \lambda }\right)+\nonumber\\
   &\hspace{1mm}2 \eta ^2 \lambda ^2\Big)-\log \left(a_1\right) (2 \eta  \lambda +1)-2
   \log ^2\left(a_1\right)\Big)-2 a_1 a_2 b_1 c_5 \kappa  e^{\eta  \lambda }\Big)\Bigg),
   \end{align}
   
\begin{align}
   \mathcal{H}_3(\eta)=&\hspace{1mm}2 e^{2 \eta  \lambda }-\frac{1}{8 a_1 a_2^2 \left(a_1 e^{2 \eta  \lambda }-a_2\right)}a \Bigg[8 a_2 a_1^3 \lambda  e^{2 \eta  \lambda } \Big(4 a_2 \kappa  e^{2 \eta 
   \lambda } \left(a_4 \gamma +a_6 \kappa +a_3\right)-2 a_2^2+\nonumber\\
   &\hspace{1mm}c_1^2 \kappa  e^{2 \eta  \lambda } \left(2 \eta  \lambda -\log
   \left(a_2-a_1 e^{2 \eta  \lambda }\right)\right)\Big)+a_1^2 \kappa  e^{2 \eta  \lambda } \Big(c_1^4
   \kappa  \lambda  e^{2 \eta  \lambda } \Big(\log \left(a_2-a_1 e^{2 \eta  \lambda }\right)\nonumber\\
   &\hspace{1mm}-2 \eta  \lambda
   \Big){}^2+8 a_2 a_3 c_1^2 \kappa  \lambda  e^{2 \eta  \lambda } \left(2 \eta  \lambda -\log \left(a_2-a_1 e^{2
   \eta  \lambda }\right)\right)+8 a_2^2 \Big(2 a_3^2 \kappa  \lambda  e^{2 \eta  \lambda }-\nonumber\\
   &\hspace{1mm}c_1 \left(c_1 \lambda 
   \left(-\log \left(a_2-a_1 e^{2 \eta  \lambda }\right)+2 \eta  \lambda +1\right)+2 c_5 \kappa \right)\Big)-32
   a_2^3 \lambda  \big(a_4 \gamma +a_6 \kappa +\nonumber\\
   &\hspace{1mm} a_3\big)\Big)-a_2 a_1 \kappa ^2 \Big(16 a_2^2 \left(a_3^2
   \lambda  e^{2 \eta  \lambda }-c_1 c_5\right)+c_1^4 \lambda  e^{2 \eta  \lambda } \Big(\log \left(a_2-a_1 e^{2
   \eta  \lambda }\right)^2-\nonumber\\
   &\hspace{1mm}2 (2 \eta  \lambda +1) \log \left(a_2-a_1 e^{2 \eta  \lambda }\right)+4 \eta  \lambda 
   (\eta  \lambda +1)\Big)+8 a_2 a_3 c_1^2 \lambda  e^{2 \eta  \lambda } \Big(2 \eta  \lambda -\nonumber\\
   &\hspace{1mm}\log \left(a_2-a_1
   e^{2 \eta  \lambda }\right)\Big)\Big)-8 a_2^3 a_3 c_1^2 \kappa ^2 \lambda +16 a_2^2 a_1^4 \lambda  e^{4 \eta 
   \lambda }\Bigg],\\
   \mathcal{H}_4(\eta)=&\hspace{1mm}\frac{a}{\lambda},
    \end{align}
   
\begin{align}
   \mathcal{H}_5(\eta)=&\hspace{1mm} a \Bigg[\frac{1}{4} \kappa  e^{\eta  \lambda } \Bigg(\frac{2 b_1 c_1^2 e^{2 \eta  \lambda }}{a_2 \left(a_2-a_1
   e^{2 \eta  \lambda }\right)}+\frac{b_1 }{a_1 a_2}\Big(c_1^2 \left(-\log \left(a_2-a_1 e^{2 \eta  \lambda }\right)+2 \eta 
   \lambda +2\right)+\nonumber\\
   &\hspace{1mm}4 a_2 a_3\Big)-\frac{d_1^2 e^{\eta  \lambda }}{a_1 a_2 \lambda  e^{2 \eta  \lambda
   }+a_2^2 \lambda }-\frac{d_1^2 \tan ^{-1}\left(\frac{\sqrt{a_1} e^{\eta  \lambda }}{\sqrt{a_2}}\right)}{\sqrt{a_1}
   a_2^{3/2} \lambda }-\frac{a_1 e_1^2 e^{\eta  \lambda }}{a_1 \lambda  e^{2 \eta  \lambda }+a_2 \lambda
   }-\nonumber\\
   &\hspace{1mm}\frac{\sqrt{a_1}}{\sqrt{a_2}
   \lambda } e_1^2 \tan ^{-1}\left(\frac{\sqrt{a_1} e^{\eta  \lambda }}{\sqrt{a_2}}\right)\Bigg)+\frac{\kappa ^2 e^{-\eta  \lambda } }{2 a_1^2 \lambda }\Big(2 a_1^2 b_4 \lambda  e^{2 \eta  \lambda }+2 a_2 a_3
   \lambda  (2 \eta  \lambda +1)-\nonumber\\
   &\hspace{1mm}b_1 c_1 c_5 (2 \eta  \lambda +1)\Big)+b_2 \gamma  \kappa 
   e^{\eta  \lambda }+b_1 e^{\eta  \lambda }\Bigg],
\end{align}
where we denote the constant $a= \frac{2 \left(a_1^2+1\right)}{3 a_1^2 \lambda }-\frac{4 a_3 \kappa
   }{3 a_1^3 \lambda }$.

   \section{Order $\kappa^2$ solutions}\label{jhepappendixk2}
In this Appendix, we note down the solution of the equations (\ref{bhjhep1eq})-(\ref{bhjhepfeq}),
\begin{align}
    \omega_3^{(bh)}=&\hspace{1mm}\frac{1}{192 \Lambda ^2 \mu ^{3/2} \text{$\phi_0$}^4}\Bigg(64 g_1 \Lambda  \mu ^{3/2} m_1^2 \rho  \text{$\phi_0$}
   \left(\log \left(1-\frac{\rho }{\sqrt{\mu }}\right)+\log
   \left(\frac{\rho }{\sqrt{\mu }}+1\right)\right)-\nonumber\\
   &\hspace{1mm}16 \Lambda
    l_1^2 m_1^2 \text{$\phi_0 $} \Big(\sqrt{\mu } \log
   \left(\frac{\rho }{\sqrt{\mu }}+1\right)+\log
   \left(1-\frac{\rho }{\sqrt{\mu }}\right) \Big(\rho
   \left(-\log \left(\frac{\rho }{\sqrt{\mu
   }}+1\right)\right)+\nonumber\\
   &\hspace{1mm}\sqrt{\mu }+\rho  \log (4)\Big)+\rho
   \tanh ^{-1}\left(\frac{\rho }{\sqrt{\mu }}\right)-2 \rho
   \text{Li}_2\left(\frac{1}{2}-\frac{\rho }{2 \sqrt{\mu
   }}\right)+\rho \Big)+16 m_1 \times\nonumber\\
   &\hspace{1mm}\Big(3 \mu  \rho  \tanh
   ^{-1}\left(\frac{\rho }{\sqrt{\mu }}\right) \left(2
   \Lambda ^2 m_6 \text{$\phi_0$}^3-3 m_1^3\right)-\Lambda
   m_1 n_1^2 \text{$\phi_0$} \Big(\sqrt{\mu } \log
   \left(\frac{\rho }{\sqrt{\mu }}+1\right)\nonumber\\
   &\hspace{1mm}+\log
   \left(1-\frac{\rho }{\sqrt{\mu }}\right) \left(\rho
   \left(-\log \left(\frac{\rho }{\sqrt{\mu
   }}+1\right)\right)+\sqrt{\mu }+\rho  \log (4)\right)+\rho
   \tanh ^{-1}\left(\frac{\rho }{\sqrt{\mu }}\right)\nonumber\\
   &\hspace{1mm}-2 \rho
   \text{Li}_2\left(\frac{1}{2}-\frac{\rho }{2 \sqrt{\mu
   }}\right)+\rho \Big)\Big)-96 \Lambda  \sqrt{\mu } q_2
   \text{$\phi_0$}^2 \Big(\Lambda  q_1 \text{$\phi_0$}^2
   \Big(-\sqrt{\mu }+2 \left(\mu -\rho ^2\right)
   \rho\times \nonumber\\
   &\hspace{1mm} \tanh
   ^{-1}\left(\frac{\rho }{\sqrt{\mu }}\right)\Big)-4 \sqrt{\mu } m_1^2 \rho  \tanh
   ^{-1}\left(\frac{\rho }{\sqrt{\mu }}\right)\Big)-3
   \Lambda ^2 \sqrt{\mu } q_2^2 \text{$\phi_0$}^4 \Big(
   \tanh ^{-1}\left(\frac{\rho }{\sqrt{\mu }}\right)\nonumber\\
   &\hspace{1mm}\times32\left(\left(\mu -\rho ^2\right) \tanh
   ^{-1}\left(\frac{\rho }{\sqrt{\mu }}\right)+\sqrt{\mu }
   \rho \right)+45 \sqrt{\mu } \rho \Big)+96 \Lambda ^2
   \sqrt{\mu } \rho  q_1^2 \text{$\phi_0$}^4 \Big(\rho
   -\nonumber\\
   &\hspace{1mm}\sqrt{\mu } \tanh ^{-1}\left(\frac{\rho }{\sqrt{\mu
   }}\right)\Big)+192 \Lambda ^2 \mu  \text{$\phi_0$}^4
   \left(q_6 \left(\rho  \tanh ^{-1}\left(\frac{\rho
   }{\sqrt{\mu }}\right)-\sqrt{\mu }\right)+\rho
   q_5\right)\Bigg),\label{jhepeopmkappa2i}\\
   \Phi_3^{(bh)}=&\hspace{1mm}g_3\rho-\frac{1}{8 \mu ^2}\Bigg(2 \rho ^2 \tanh ^{-1}\left(\frac{\rho }{\sqrt{\mu
   }}\right) \left(q_1 \left(l_1^2+n_1^2\right)-4 g_1 \mu
   ^{3/2} q_2\right)+\mu  \log \left(\rho -\sqrt{\mu }\right)\times\nonumber\\
   &\hspace{1mm}\left(q_1 \left(l_1^2+n_1^2\right)-4 g_1 \mu ^{3/2}
   q_2\right)-\mu  \log \left(\sqrt{\mu }+\rho \right)
   \left(q_1 \left(l_1^2+n_1^2\right)-4 g_1 \mu ^{3/2}
   q_2\right)-\nonumber\\
   &\hspace{1mm}8 g_1 \mu ^{3/2} \rho ^2 q_1-8 g_1 \mu ^2 \rho
   q_2+2 \mu  q_2 \left(l_1^2+n_1^2\right) \log
   \left(1-\frac{\rho ^2}{\mu }\right)-\mu  \big(q_2
   \left(l_1^2+n_1^2\right)+\nonumber\\
   &\hspace{1mm}4 l_1 l_5+2 n_1 n_5\big) \log
   \left(\mu -\rho ^2\right)+2 q_2 \left(l_1^2+n_1^2\right)
   \left(\rho ^2-\mu \right) \tanh ^{-1}\left(\frac{\rho
   }{\sqrt{\mu }}\right)^2+\nonumber\\
   &\hspace{1mm}2 \sqrt{\mu } \rho  q_1
   \left(l_1^2+n_1^2\right)-\mu  \left(q_2
   \left(l_1^2+n_1^2\right)-4 l_1 l_5-2 n_1 n_5\right) \log
   \left(\sqrt{\mu }+\rho \right)+\nonumber\\
   &\hspace{1mm}\sqrt{\mu } \left(\rho
   -\sqrt{\mu }\right) \left(q_2 \left(l_1^2+n_1^2\right)-4
   l_1 l_5-2 n_1 n_5\right) \log \left(\rho -\sqrt{\mu
   }\right)-\sqrt{\mu } \rho  \big(q_2
   \left(l_1^2+n_1^2\right)\nonumber\\
   &\hspace{1mm}-4 l_1 l_5-2 n_1 n_5\big) \log
   \left(\sqrt{\mu }+\rho \right)+4 \sqrt{\mu } \rho  q_2
   \left(l_1^2+n_1^2\right) \tanh ^{-1}\left(\frac{\rho
   }{\sqrt{\mu }}\right)\Bigg)+g_4,\label{jhepeopmkappa2i1}
   \end{align}
   \begin{align}
   A_{t3}^{(bh)}=&\hspace{1mm}\frac{1}{8 \Lambda  \mu ^{3/2}
   \text{$\phi_0$}^2}\Bigg(2 \sqrt{\mu } \Big(m_1 \rho  \left(-4 g_1 \mu  \rho
   +l_1^2+n_1^2+8 \sqrt{\mu } \rho  q_1 \text{$\phi_0$}+8 \mu
    q_2 \text{$\phi_0$}\right)+\nonumber\\
    &\hspace{1mm}4 \Lambda  \mu  \text{$\phi
_0$}^2 \left(m_6 \rho +m_5\right)\Big)+2 m_1 \rho ^2
   \tanh ^{-1}\left(\frac{\rho }{\sqrt{\mu }}\right)
   \left(l_1^2+n_1^2+8 \mu  q_2 \text{$\phi_0$}\right)+\nonumber\\
   &\hspace{1mm}\mu
   m_1 \left(\log \left(\rho -\sqrt{\mu }\right)-\log
   \left(\sqrt{\mu }+\rho \right)\right) \left(l_1^2+n_1^2+8
   \mu  q_2 \text{$\phi_0$}\right)\Bigg),\\
   \xi_3^{(bh)}=&\hspace{1mm}\frac{l_5}{\sqrt{\mu}}\tanh ^{-1}\left(\frac{\rho }{\sqrt{\mu }}\right)+l_6,\label{jhepeopmkappa2i2}\\
   \chi_3^{(bh)}=&\hspace{1mm}n_5t+n_6,\label{jhepeopmkappa2f}
\end{align}
where $m_i,n_i,l_i,q_i$, $g_j$, $(i=5,6$, $j=3,4)$ are the integration constants and we define $\text{Li}_2(x)=\text{PolyLog}(2,x)$.

%%%%%%%%%%%%%%%%%%%%%%%%%%%%%

\end{document}